\documentclass[11pt,english]{article}
\usepackage[latin9]{inputenc}
\usepackage{amsmath}
\usepackage{amssymb}
\usepackage{cancel}
\usepackage{graphicx}
\usepackage{setspace}
\makeatletter

\providecommand{\tabularnewline}{\\}

\@ifundefined{date}{}{\date{}}
\usepackage{esint}
\usepackage{hyperref}
\usepackage{latexsym}
\usepackage{graphicx}\usepackage{bm}\usepackage{longtable}

\usepackage{xcolor}

\@addtoreset{equation}{section}

\makeatother

\usepackage{babel}
\begin{document}
\title{Holographic spectroscopy of fermion with instantons}
\maketitle
\begin{center}
Si-wen Li\footnote{Email: siwenli@dlmu.edu.cn}, Yi-peng Zhang\footnote{Email: ypmahler111@dlmu.edu.cn},
Hao-qian Li\footnote{Email: lihaoqian@dlmu.edu.cn}, 
\par\end{center}

\begin{center}
\emph{Department of Physics, School of Science,}\\
\emph{Dalian Maritime University, }\\
\emph{Dalian 116026, China}\\
\par\end{center}

\vspace{12mm}

\begin{abstract}
Using the gauge-gravity duality, we investigate the fermionic spectroscopy
in the D(-1)-D3 brane system. The background geometry of this system
described by IIB supergravity includes a black (deconfined) and bubble
(confined) D3-brane which correspond respectively to a deconfined
and a confined gauge theory in holography. The charge of the D(-1)
brane as the D-instanton gives the gluon condensate in this model.
To simplify the holographic setup, we first reduce briefly the ten-dimensional
supergravity background produced by D(-1)-D3-branes to an equivalent
five-dimensional background. Then the fermionic spectrum in the confined
case is obtained by decomposing the fermion using dimensional reduction.
In addition, by using the standard method for computing the Green's
function in the AdS/CFT dictionary, we derive the equations for the
fermionic correlation functions and solve them numerically with the
infalling boundary condition. Our numerical results in the deconfined
case illustrate that the fermionic correlation function as spectral
function includes two branches of the dispersion curves whose behavior
is very close to the results obtained from the method of hard thermal
loop. And the effective mass generated by the medium effect of fermion
splits into two values due to the spin-dependent interactions induced
by instantons. In the confined case, the holographic correlation function
indicates several separated dispersion curves which illustrates consistently
the onset mass in the fermionic spectrum we obtained. Therefore, this
work on holography demonstrates the instantonic configuration is very
influential to the fermion in QCD.
\end{abstract}
\newpage{}

\tableofcontents{}

\section{Introduction}

In quantum chromodynamics (QCD), the instanton is known as the non-trivially
topological excitation of the vacuum which contributes to the thermodynamics
of QCD and relates to the breaking of chiral symmetry \cite{key-1,key-2,key-3}.
The instanton is also known to consist of constituents which are called
BPS (Bogomol'nyi-Prasad-Sommerfield) monopoles or dyons \cite{key-4},
so there have been a lot of continuous researches to connect confinement
and the breaking of chiral symmetry to the instanton constituents
\cite{key-5,key-6,key-7,key-8,key-9,key-10,key-11,key-12,key-13}.
In particular, the dynamics of fermions involving instantons e.g.
the spin-dependent interaction, fermionic quantum tunneling of the
instantonic vacuum, is very interesting and widely discussed in the
textbooks of QCD \cite{key-1,key-2,key-3}. 

However, due to the property of asymptotic freedom, it is very challenging
to investigate QCD in the low-energy region by using the perturbative
method in quantum field theory (QFT) since QCD is strongly coupled
in the low-energy region. Fortunately, the gauge-gravity duality provides
an alternative way to study the strongly coupled QFT \cite{key-14,key-15,key-16}
using holography. In the top-down approach, the instanton is constructed
simply as a geometric background with instantonic D-branes (D-instanton)
\cite{key-17,key-18,key-19,key-20,key-21}, for example, the black
D(-1)-D3 brane system can describe holographically the plasma with
instantons in which the D(-1)-brane plays the role of instanton, and
its dual field theory is the $\mathcal{N}=4$ super Yang-Mills theory
with an instanton or a theta angle. So there have been many works
based on the holographic instanton e.g. thermodynamics of Yang-Mills
theory \cite{key-22}, hadronic spectrum \cite{key-23,key-24}, real-time
dynamics \cite{key-25}, Chern-Simons theory \cite{key-26,key-27},
chiral symmetry breaking \cite{key-28}, quark potential \cite{key-29,key-30,key-31},
Schwinger effect \cite{key-32,key-33,key-34}, the hadronic interaction
\cite{key-36}. Despite all this, the study of fermions in the presence
of the instanton remains unexplored in the existing works in holography
although it attracts great interests as it is reviewed in \cite{key-1,key-2,key-3}
in QFT. Therefore, the goal of this work is to explore the dynamics
of fermions in the presence of an instanton as a starting point for
further investigations, since it can be evaluated by using the most
fundamental principle in holography.

For our goal, the setup in this work starts with the D(-1)-D3 brane
system since it describes simply the instantons in gauge theory through
gauge-gravity duality \cite{key-17,key-25,key-28}. Then, we reduce
the 10d IIB supergravity background produced by black D3-branes with
D(-1)-branes to an effective five-dimensional (5d) geometric background.
We note that the supergravity solution for the D(-1)-D3 brane system
includes a black (deconfined) and a bubble (confined) D3-brane solution
which corresponds respectively to a deconfined and a confined gauge
theories with instantons in holography \cite{key-a1,key-a2}. Afterwards,
we study a probe bulk fermion in the 5d geometric background and derive
its covariant Dirac equation. In the confined geometry, we decompose
the bulk fermion using dimensional reduction, evaluate numerically
the holographic spectrum of fermion. Since the bulk fermion must be
a gauge-invariant operator, we identify it with a baryon in field
theory\footnote{Since baryon is the fermionic gauge-invariant operator in QCD, for
a bottom-up or phenomenological approach of QCD, the fermionic gauge-invariant
operator is usually identified to a baryon, e.g. \cite{key-b3}. The
top-down approaches also support that the gauge-invariant fermions
behave as baryons as it is discussed in \cite{key-b8,key-b6}. In
particular, baryon can be introduced into the holographic approaches
by considering the baryon vertex as it is discussed in \cite{key-b4}
and the relevant works with baryon vertex can also be reviewed in
\cite{key-36,key-35,key-b5}.}. In addition, we investigate the fermionic two-point correlation
function in the dual theory by employing the standard technique in
the AdS/CFT dictionary so that obtain the differential equations for
the fermionic correlators in the bulk\footnote{The two-point functions of fermions in the context of the AdS/CFT
were first considered in \cite{key-b2,key-b7}}. By imposing the infalling boundary condition at the horizon, we
afterwards solve the fermionic correlation functions numerically both
in deconfined and confined backgrounds. Our numerical calculation
in the deconfined case illustrates that, without the D-instanton,
there is only one branch of the holographic dispersion curves for
fermion lying on the light-cone which is similar as the existing works
about holographic fermions \cite{key-36,key-35,key-37,key-38,key-39,key-40}.
In the presence of the D-instanton, the fermionic correlation function
has a correction which to leading order splits the dispersion curves
into two branches at small momentum, and this behavior is very close
to the dispersion curves from the approach of hard thermal loop (HTL)
in QFT. Furthermore, the two branches of the dispersion curves reveals
that the effective bound mass of fermion in plasma has two values
in the presence of the D-instanton. Remarkably it agrees qualitatively
with effective potential for the spin-dependent interactions induced
by instantons computed by the process of one-gluon exchange in QCD
\cite{key-1,key-2}. In the confined case, the holographic correlation
function indicates several separated dispersion curves which agrees
with the fermionic spectrum we obtained. Therefore, this work, as
a starting point, is a holographic reproduction of the properties
of fermions in the presence of instantons, we believe the instantonic
configuration in QCD is very influential to the fermion.

The outline of this manuscript is as follows. In Section 2, we collect
the essentials of the 10d supergravity background produced by D3-branes
with D(-1)-branes as D-instantons, and reduce it briefly to an effectively
5d gravity background. In Section 3, we decompose the bulk spinor
with dimensional reduction, then evaluate the eigenmass of fermion
numerically. In Section 4, we demonstrate the principle in AdS/CFT
dictionary to compute the holographic correlation function for spinor.
Then we derive the covariant Dirac equation and the holographic counter
term for spinor in order to obtain the differential equations for
the fermionic correlators. In Section 4, we solve numerically the
holographic fermionic correlation functions order by order with physical
analysis and interpretation, both in the deconfined and confined case.
The final section is the summary. 

\subsection*{Notation}

In this manuscript, the capital letters $L,M,N...$ refer to the indices
of the bulk coordinate as $x^{M}$, the lowercase letters $a,b,c$
refer to the corresponding indices of the coordinate in tangent space.
The gamma matrices $\gamma^{a},\Gamma^{M}$ as the generators of the
Clifford algebra with vielbein $e_{M}^{a}$ are related as,

\begin{equation}
\left\{ \gamma^{a},\gamma^{b}\right\} =2\eta^{ab},\left\{ \Gamma^{M},\Gamma^{M}\right\} =2g^{MN},\Gamma^{M}=e_{a}^{M}\gamma^{a},g_{MN}=e_{M}^{a}\eta_{ab}e_{N}^{a},
\end{equation}
where $g_{MN}$ refers to the metric on a manifold and $\eta_{ab}$
refers to the Minkowskian metric. The spin connection $\omega_{Mab}$
is defined as,

\begin{equation}
\omega_{Mab}=\eta_{cb}\left(e_{a}^{N}\partial_{M}e_{N}^{c}-e_{a}^{N}\Gamma_{MN}^{K}e_{K}^{c}\right),
\end{equation}
where the affine connection $\Gamma_{MN}^{K}$ is given by,

\begin{equation}
\Gamma_{MN}^{K}=g^{KL}\left(\partial_{M}g_{LN}+\partial_{N}g_{ML}-\partial_{L}g_{MN}\right).
\end{equation}
The Greek letters as $\mu,\nu...$refer to the indices of coordinate
as $x^{\mu}$ at the holographic boundary. The lowercase letters $i,j,k,l$
run over the spatial directions of the holographic boundary i.e. $i,j,k,l=1,2,3$.
The gamma matrices presented in this work are chosen as $\gamma^{a}=\left\{ \gamma^{\mu},\gamma\right\} $
as,
\begin{equation}
\gamma^{\mu}=i\left(\begin{array}{cc}
0 & \sigma^{\mu}\\
\bar{\sigma}^{\mu} & 0
\end{array}\right),\gamma=\left(\begin{array}{cc}
1 & 0\\
0 & -1
\end{array}\right),
\end{equation}
with $\sigma^{\mu}=\left(1,-\sigma^{i}\right),\bar{\sigma}^{\mu}=\left(1,\sigma^{i}\right),i=1,2,3$,
where $\sigma^{i}$ s are the Pauli matrices.

\section{The instantonic background in holography }

In this section, let us briefly review the instantonic background
which corresponds to the system consisting of $N_{c}$ D3-branes with
$N_{\mathrm{D}}$ D-instantons i.e. the D(-1)-branes in the large-$N_{c}$
limit in the type IIB string theory \cite{key-17,key-18,key-25}.
Then, we integrate out the $S^{5}$ part in order to obtain a five-dimensional
(5d) background geometry for simplification. 

\subsection{The D3-brane background with D-instanton}

The supergravity solution of $N_{c}$ D3-branes with $N_{\mathrm{D}}$
D-instantons is in general described by a 10d deformed D3-brane solution
with a non-trivial Ramond-Ramond (R-R) zero form $C_{0}$. It is recognized
as a marginal \textquotedblleft bound state\textquotedblright{} of
D3-branes with smeared $N_{\mathrm{D}}$ D(-1)-branes. In our notation,
$N_{c}$ D3-branes are identified as color branes, hence the associated
number $N_{c}$ denotes the color number. In the large-$N_{c}$ limit,
the 10d bosonic type IIB supergravity action describes the low-energy
dynamics of this system which is given in the string frame as, 

\begin{equation}
S_{\mathrm{IIB}}=\frac{1}{2\kappa_{10}^{2}}\int d^{10}x\sqrt{-g}\left[e^{-2\Phi}\left(\mathcal{R}+4\partial\Phi\cdot\partial\Phi\right)-\frac{1}{2}\left|F_{1}\right|^{2}-\frac{1}{2}\left|F_{5}\right|^{2}\right],\label{eq:5}
\end{equation}
where $2\kappa_{10}^{2}=\left(2\pi\right)^{7}l_{s}^{8}$ is the 10d
gravity coupling constant, $l_{s},g_{s}$ is respectively the string
length and the string coupling constant. We use $\Phi$ to denote
the dilaton field and $F_{1,5}$ is the field strength of the R-R
zero and four form $C_{0,4}$ respectively. The solution of $N_{c}$
D3-branes with $N_{\mathrm{D}}$ D-instantons is the near-horizon
solution of non-extremal D3-branes with a non-trivial $C_{0}$, which
in string frame reads \cite{key-17,key-18,key-25},

\begin{align}
ds^{2} & =e^{\frac{\phi}{2}}\left\{ \frac{r^{2}}{R^{2}}\left[-f\left(r\right)dt^{2}+d\mathbf{x}\cdot d\mathbf{x}\right]+\frac{1}{f\left(r\right)}\frac{R^{2}}{r^{2}}dr^{2}+R^{2}d\Omega_{5}^{2}\right\} ,\nonumber \\
e^{\phi} & =1+\frac{Q}{r_{H}^{4}}\ln\frac{1}{f\left(r\right)},\ f\left(r\right)=1-\frac{r_{H}^{4}}{r^{4}},\ F_{5}=dC_{4}=g_{s}^{-1}\mathcal{Q}_{3}\epsilon_{5},\nonumber \\
F_{1} & =dC_{0},\ C_{0}=-ie^{-\phi}+i\mathcal{C},\ \phi=\Phi-\Phi_{0},\ e^{\Phi_{0}}=g_{s},\label{eq:6}
\end{align}
where $\mathcal{C}$ is a boundary constant for $C_{0}$, $\epsilon_{5}$
is the volume element of a unit $S^{5}$. And the associated parameters
are given as,

\begin{equation}
R^{4}=4\pi g_{s}N_{c}l_{s}^{4},\ \mathcal{Q}_{3}=4R^{4},\ Q=\frac{N_{\mathrm{D}}}{N_{c}}\frac{\left(2\pi\right)^{4}\alpha^{\prime2}}{V_{4}}\mathcal{Q}_{3}.
\end{equation}
Here, $x^{\mu}=\left\{ t,\mathbf{x}\right\} =\left\{ t,x^{i}\right\} ,\ i=1,2,3$
refers to the 4d spacetime where the D3-branes are extended along.
And $r$ is the holographic direction perpendicular to the D3-branes.
The solution (\ref{eq:6}) is asymptotically $\mathrm{AdS}_{5}\times S^{5}$
at the holographic boundary $r\rightarrow\infty$ which describes
geometrically that the D-instanton charge $N_{\mathrm{D}}$ is smeared
over the worldvolume $V_{4}$ of the $N_{c}$ coincident black D3-branes
homogeneously with a horizon at $r=r_{H}$\footnote{When the IR cut-off is taken into account, $V_{4}$ can be chosen
to be finite \cite{key-17}. }. The backreaction of the D-instantons has been taken into account
in the background, thus it implies $N_{\mathrm{D}}/N_{c}$ must be
fixed in the large-$N_{c}$ limit. The dual theory of this background
is conjectured to be the 4d $\mathcal{N}=4$ super Yang-Mills theory
(SYM) in a self-dual gauge field (instantonic) background or with
a dynamical axion at finite temperature characterized by the parameter
$Q$ \cite{key-17,key-18}. Note that, in the language of hadron physics,
$C_{0}$ is recognized as the axion. The gluon condensate in this
system can be evaluated as,

\begin{equation}
\left\langle \mathrm{Tr}F_{\mu\nu}F^{\mu\nu}\right\rangle \simeq\frac{N_{\mathrm{D}}}{16\pi^{2}V_{4}}=\frac{1}{16}\frac{Q}{\left(2\pi\alpha^{\prime}\right)^{2}R^{4}}\frac{N_{c}}{\left(2\pi\right)^{4}},\ \mu,\nu=0,1...3.\label{eq:8}
\end{equation}

Besides, by following \cite{key-a3,key-a4}, there is a confining
background which can be obtained from (\ref{eq:6}). That is to compactify
one of the spatial dimension $x^{3}=y$ of the D3-brane on a circle
$S^{1}$, so the worldvolume theory on the D3-branes is effectively
3d below the energy scale $M_{KK}=\frac{2\pi}{\delta y}$, where $\delta y$
refers to the size of $S^{1}$. Then identify the bulk supergravity
geometry that has this geometry for its boundary, which resultantly
leads to a double Wick rotation ($t\rightarrow-iy,y\rightarrow-it$)
on the background presented in (\ref{eq:6}) as,

\begin{align}
ds^{2} & =e^{\frac{\phi}{2}}\left\{ \frac{r^{2}}{R^{2}}\left[\eta_{\mu\nu}dx^{\mu}dx^{\nu}+f\left(r\right)dy^{2}\right]+\frac{1}{f\left(r\right)}\frac{R^{2}}{r^{2}}dr^{2}+R^{2}d\Omega_{5}^{2}\right\} ,\mu,\nu=0,1,2.\nonumber \\
f\left(r\right) & =1-\frac{r_{KK}^{4}}{r^{4}}.\label{eq:9}
\end{align}
The background (\ref{eq:9}) is defined only for $r>r_{KK}$ and it
does not have a horizon, so $r_{KK}$ is the end of the spacetime.
Since the warp factor $e^{\frac{\phi}{2}}\frac{r^{2}}{R^{2}}$ never
goes to zero, the Wilson loop asymptotics lead to an area law which
manifests as confinement in the dual theory.

\subsection{The effective 5d background}

The $S^{5}$ part in the gravity solution (\ref{eq:5}) and (\ref{eq:6})
can be integrated out so that the background (\ref{eq:6}) can reduce
to an equivalent 5d background. Since all the fields presented in
(\ref{eq:5}) and (\ref{eq:6}) are expected to depend on the coordinates
$x^{\mu},r$ only, it is possible to integrate out the $S^{5}$ part
in action with

\begin{equation}
ds_{\left(10\mathrm{d}\right)}^{2}=ds_{\left(5\mathrm{d}\right)}^{2}+ds_{S^{5}}^{2}.
\end{equation}
So in the Einstein frame, the action (\ref{eq:5}) becomes,

\begin{equation}
S_{\mathrm{IIB}}^{\left(5\mathrm{d}\right)}=\frac{1}{2\kappa_{5}^{2}}\int d^{5}x\sqrt{-g^{\left(5\mathrm{d}\right)}}\left[\mathcal{R}^{5\mathrm{d}}-\frac{1}{2}\partial\Phi\cdot\partial\Phi-\frac{1}{2}\left|dC_{0}\right|^{2}-2\Lambda\right],\label{eq:11}
\end{equation}
where the cosmological constant $\Lambda$ is obtained by

\begin{equation}
\Lambda=\frac{\left(2\pi l_{s}\right)^{5}}{2}\int_{S^{5}}dx^{5}\sqrt{-g_{S^{5}}}\left|F_{5}\right|^{2}=-\frac{6}{R^{2}}.
\end{equation}
Since the supergravity solution (\ref{eq:6}) implies that the kinetic
terms of $\Phi,C_{0}$ presented in (\ref{eq:11}) cancel each other,
the 5d solution is nothing but $\mathrm{AdS_{5}}$. Thus, in the string
frame it is given as,

\begin{equation}
ds_{\left(5\mathrm{d}\right)}^{2}=e^{\frac{\phi}{2}}\left\{ \frac{r^{2}}{R^{2}}\left[-f\left(r\right)dt^{2}+d\mathbf{x}\cdot d\mathbf{x}\right]+\frac{1}{f\left(r\right)}\frac{R^{2}}{r^{2}}dr^{2}\right\} ,\label{eq:13}
\end{equation}
and the solution for $\Phi$ and $C_{0}$ are still given by the corresponding
formulas given in (\ref{eq:6}). Note that the action (\ref{eq:11})
describes nothing but the gravity-dilaton-axion system, and we will
use the 5d effective background (\ref{eq:13}) in $z$ coordinate
as,

\begin{align}
ds_{\left(5\mathrm{d}\right)}^{2}= & e^{\frac{\phi}{2}}\frac{R^{2}}{z^{2}}\left[-f\left(z\right)dt^{2}+d\mathbf{x}\cdot d\mathbf{x}+\frac{dz^{2}}{f\left(z\right)}\right],\nonumber \\
f\left(z\right)= & 1-\frac{z^{4}}{z_{H}^{4}},e^{\phi}=1-q\ln f\left(z\right),\nonumber \\
z= & \frac{R^{2}}{r},z_{H}=\frac{R^{2}}{r_{H}},q=\frac{z_{H}^{4}Q}{R^{8}}.\label{eq:14}
\end{align}
in which the holographic boundary is located at $z\rightarrow0$ to
continue our discussion. Moreover, the confined geometry presented
(\ref{eq:9}) also reduces to an effective 5d background as,

\begin{align}
ds_{\left(5\mathrm{d}\right)}^{2}= & e^{\frac{\phi}{2}}\frac{R^{2}}{z^{2}}\left[\eta_{\mu\nu}dx^{\mu}dx^{\nu}+f\left(r\right)dy^{2}+\frac{dz^{2}}{f\left(z\right)}\right],\nonumber \\
f\left(z\right)= & 1-\frac{z^{4}}{z_{KK}^{4}},e^{\phi}=1-q\ln f\left(z\right),\nonumber \\
z= & \frac{R^{2}}{r},z_{H}=\frac{R^{2}}{r_{KK}},q=\frac{z_{KK}^{4}Q}{R^{8}}.\label{eq:15}
\end{align}

\section{The spectrum of the baryonic fermion}

\begin{figure}
\begin{centering}
\includegraphics[scale=0.27]{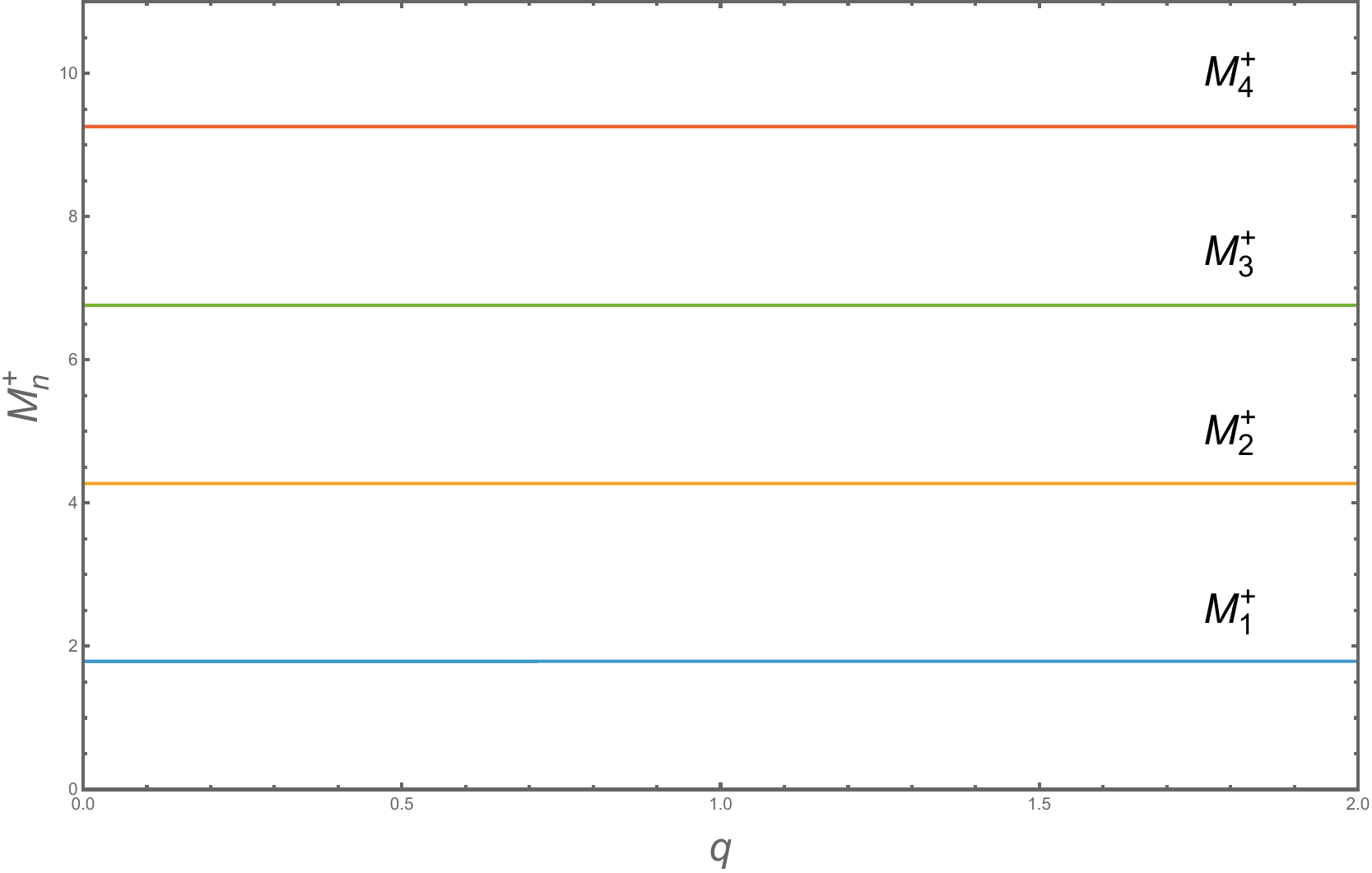}\includegraphics[scale=0.27]{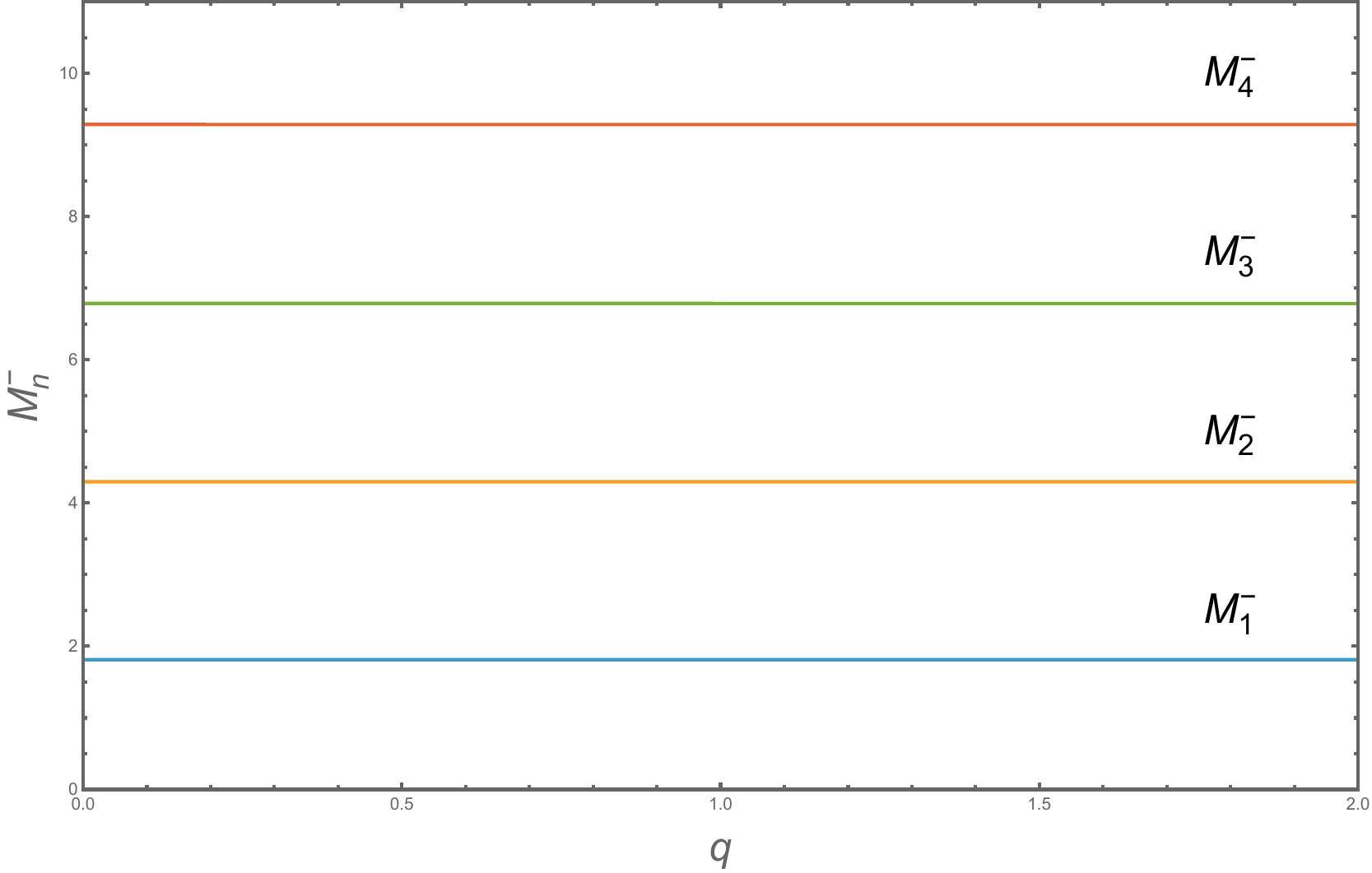}
\par\end{centering}
\caption{\label{fig:1}The mass spectrum as a function of the instanton density
$q$.}

\end{figure}

In this section, our concern is the fermionic spectrum in the confined
geometry. As the Wilson loop asymptotics lead to an area law, we consider
a fermionic operator in the boundary theory which describes a confined
state. In particular, the fermionic operator should be gauge-invariant
according to gauge-gravity duality \footnote{The gauge-invariant fermion in the confined phase of QCD is a baryon.
In the top-down approach, it can be constructed by introducing a baryon
vertex \cite{key-35,key-36,key-b5}. However, since we are following
a bottom-up approach, we will not attempt to discuss the holographic
construction of a baryon with a baryon vertex in this work.}\cite{key-15,key-16,key-a4}. As the confined metric given in (\ref{eq:15})
becomes 3d below the energy scale $M_{KK}$, it reduces to the following
metric,

\begin{align}
ds_{\left(4\mathrm{d}\right)}^{2} & =e^{\frac{\phi}{2}}\frac{R^{2}}{z^{2}}\left[\eta_{\mu\nu}dx^{\mu}dx^{\nu}+\frac{dz^{2}}{f\left(z\right)}\right],\nonumber \\
 & \equiv g_{xx}\eta_{\mu\nu}dx^{\mu}dx^{\nu}+g_{zz}dz^{2}.\label{eq:16}
\end{align}
We start with a bulk fermion $\psi$ as the dual mode to the baryonic
fermion at the boundary propagating on the background (\ref{eq:16}).
Its Dirac action is given as,

\begin{equation}
S_{\mathrm{bulk}}=i\int d^{4}x\sqrt{-g}\bar{\psi}\left(\Gamma^{M}\nabla_{M}-m\right)\psi,\label{eq:17}
\end{equation}
where the Dirac operator is computed as,

\begin{equation}
\Gamma^{M}\nabla_{M}=\frac{1}{\sqrt{g_{xx}}}\gamma^{\mu}\partial_{\mu}+\frac{1}{\sqrt{g_{zz}}}\gamma\partial_{z}+\frac{3}{4}\frac{\partial_{z}\ln g_{xx}}{\sqrt{g_{zz}}}\gamma.
\end{equation}
So the Dirac equation (\ref{eq:17}) can be rewritten as,

\begin{equation}
S_{\mathrm{bulk}}=i\int d^{3}xdz\bar{\psi}\left(A_{1}\gamma^{\mu}\partial_{\mu}+A_{2}\gamma\partial_{z}+A_{3}\gamma+A_{4}\right)\psi,\label{eq:19}
\end{equation}
where the coefficients are

\begin{equation}
A_{1}=\sqrt{-\frac{g}{g_{xx}}},A_{2}=\sqrt{-\frac{g}{g_{zz}}},A_{3}=\frac{3}{4}\partial_{z}\ln g_{xx}\sqrt{-\frac{g}{g_{zz}}},A_{4}=m\sqrt{-g}.
\end{equation}
Let us further decompose the spinor $\psi$ as a series of the basis
function as,

\begin{equation}
\psi=\sum_{n,k}\left(\begin{array}{c}
\psi_{+}^{\left(n\right)}F_{+}^{\left(n\right)}\\
\psi_{-}^{\left(k\right)}F_{-}^{\left(k\right)}
\end{array}\right),\label{eq:21}
\end{equation}
where $\psi_{\pm}^{\left(k\right)}$ depends on $x^{\mu}$ and $F_{\pm}^{\left(k\right)}$
depends on $z$. Then, we insert the decomposition in the action (\ref{eq:19}),
and it becomes

\begin{align}
S_{\mathrm{bulk}}= & -\sum_{k,n}\int d^{3}xdz\bigg[\psi_{-}^{\left(k\right)\dagger}F_{-}^{\left(k\right)}A_{2}\psi_{+}^{\left(n\right)}\partial_{z}F_{+}^{\left(n\right)}+\psi_{-}^{\left(k\right)\dagger}F_{-}^{\left(k\right)}A_{3}\psi_{+}^{\left(n\right)}F_{+}^{\left(n\right)}\nonumber \\
 & +\psi_{-}^{\left(k\right)\dagger}F_{-}^{\left(k\right)}A_{4}\psi_{+}^{\left(n\right)}F_{+}^{\left(n\right)}+\psi_{-}^{\left(k\right)\dagger}F_{-}^{\left(k\right)}A_{1}i\sigma^{\mu}\partial_{\mu}\psi_{-}^{\left(n\right)}F_{-}^{\left(n\right)}\nonumber \\
 & +\psi_{+}^{\left(n\right)\dagger}F_{+}^{\left(n\right)}A_{1}i\bar{\sigma}^{\mu}\partial_{\mu}\psi_{+}^{\left(k\right)}F_{+}^{\left(k\right)}-\psi_{+}^{\left(n\right)\dagger}F_{+}^{\left(n\right)}A_{2}\psi_{-}^{\left(k\right)}\partial_{z}F_{-}^{\left(k\right)}\nonumber \\
 & -\psi_{+}^{\left(n\right)\dagger}F_{+}^{\left(n\right)}A_{3}\psi_{-}^{\left(k\right)}F_{-}^{\left(k\right)}+\psi_{+}^{\left(n\right)\dagger}F_{+}^{\left(n\right)}A_{4}\psi_{-}^{\left(k\right)}F_{-}^{\left(k\right)}\bigg].\label{eq:22}
\end{align}
By imposing the normalization condition for the basis function,

\begin{equation}
\int dzF_{+}^{\left(k\right)}F_{+}^{\left(n\right)}=\int dzF_{-}^{\left(k\right)}F_{-}^{\left(n\right)}=\delta^{kn},
\end{equation}
with eigenequations

\begin{align}
\frac{A_{2}}{\sqrt{A_{1}}}\partial_{z}\left(\frac{F_{-}^{\left(n\right)}}{\sqrt{A_{1}}}\right)+\frac{A_{3}-A_{4}}{A_{1}}F_{-}^{\left(n\right)} & =-M_{n}F_{+}^{\left(n\right)},\nonumber \\
\frac{A_{2}}{\sqrt{A_{1}}}\partial_{z}\left(\frac{F_{+}^{\left(n\right)}}{\sqrt{A_{1}}}\right)+\frac{A_{3}+A_{4}}{A_{1}}F_{+}^{\left(n\right)} & =M_{n}F_{-}^{\left(n\right)},\label{eq:24}
\end{align}
action (\ref{eq:22}) reduces to the standard form of kinetic term
for fermion as,

\begin{equation}
S=-\sum_{n}\int d^{3}x\bigg\{\psi_{-}^{\left(n\right)\dagger}i\sigma^{\mu}\partial_{\mu}\psi_{-}^{\left(n\right)}+\psi_{+}^{\left(n\right)\dagger}i\bar{\sigma}^{\mu}\partial_{\mu}\psi_{+}^{\left(n\right)}+M_{n}\left[\psi_{-}^{\left(n\right)\dagger}\psi_{+}^{\left(n\right)}+\psi_{+}^{\left(n\right)\dagger}\psi_{-}^{\left(n\right)}\right]\bigg\}.
\end{equation}
The eigenvalue $M_{n}$ can be evaluated by solving numerically the
following eigenequations

\begin{align}
 & \frac{A_{2}}{\sqrt{A_{1}}}\partial_{z}\left(\frac{A_{2}}{A_{1}}\partial_{z}\left(\frac{F_{+}^{\left(n\right)}}{\sqrt{A_{1}}}\right)+\frac{A_{3}+A_{4}}{A_{1}}F_{+}^{\left(n\right)}\right)+\frac{A_{3}-A_{4}}{A_{1}}\left(\frac{A_{2}}{\sqrt{A_{1}}}\partial_{z}\left(\frac{F_{+}^{\left(n\right)}}{\sqrt{A_{1}}}\right)+\frac{A_{3}+A_{4}}{A_{1}}F_{+}^{\left(n\right)}\right)\nonumber \\
 & =-M_{n}^{2}F_{+}^{\left(n\right)},\nonumber \\
 & \frac{A_{2}}{\sqrt{A_{1}}}\partial_{z}\left(\frac{A_{2}}{A_{1}}\partial_{z}\left(\frac{F_{-}^{\left(n\right)}}{\sqrt{A_{1}}}\right)+\frac{A_{3}-A_{4}}{A_{1}}F_{-}^{\left(n\right)}\right)+\frac{A_{3}+A_{4}}{A_{1}}\left(\frac{A_{2}}{\sqrt{A_{1}}}\partial_{z}\left(\frac{F_{-}^{\left(n\right)}}{\sqrt{A_{1}}}\right)+\frac{A_{3}-A_{4}}{A_{1}}F_{-}^{\left(n\right)}\right)\nonumber \\
 & =-M_{n}^{2}F_{-}^{\left(n\right)}.\label{eq:26}
\end{align}
We note that, the mass $m$ presented in (\ref{eq:17}) is the only
parameter to fit the experimental data. For massless fermion $m=0$,
the equations for $F_{\pm}^{\left(n\right)}$ are identical. For $m\neq0$,
the mass spectrum of $M_{n}^{\pm}$ is separated with the separation
$\sim m/2$. According to (\ref{eq:21}), the ``$\pm$'' can be
identified to the proper quantum numbers of a baryon e.g. the 3rd
component of the isospin. In this sense, we solve (\ref{eq:26}) with
sufficiently small mass parameter $m=0.01$ and the associated mass
spectrum is given in Table \ref{tab:1} 
\begin{table}
\begin{centering}
\begin{tabular}{|c|c|c|c|c|}
\hline 
$q=0$ & $n=1$ & $n=2$ & $n=3$ & $n=4$\tabularnewline
\hline 
\hline 
$M_{+}$ & 1.7899 & 4.2721 & 6.7597 & 9.2568\tabularnewline
\hline 
$M_{-}$ & 1.8131 & 4.2978 & 6.7868 & 9.2853\tabularnewline
\hline 
\end{tabular}
\par\end{centering}
\caption{\label{tab:1}The fermionic spectrum with $m=0$ in the unit of $M_{KK}=1$.}

\end{table}
 and its dependence on $q$ is illustrated in Figure \ref{fig:1}.
The numerical results are related the fact that the mass of baryons
as the up and down components of a isospin is roughly equal. The mass
spectrum is less dependent on the instanton density. However the separation
of the eigenmass does not match very well to the experimental data
for the lowest baryon mass. For example, if we identify the states
of mass $M_{1}^{\pm},M_{2}^{\pm}$ as proton ($n=1,I=+1/2$), neutron
($n=1,I=-1/2$), $\Sigma^{+}$ ($n=1,I=+1/2$), $\Sigma^{-}$ ($n=1,I=-1/2$),
the mass ratio of $\Sigma^{+}$ and proton given in this model is
$M_{2}^{+}/M_{1}^{+}\simeq2.39$ less close to the experimental data
$M_{\mathrm{proton}}/M_{\Sigma^{+}}\simeq1.27$. One possible reason
may be that the current theory is 3d instead of a 4d theory. 

\section{The fermionic correlation function}

In this section, we focus on the fermionic correlation function as
a parallel method to investigate the fermionic spectrum. By introducing
a probe fermion in the bulk geometry, we will evaluate its two-point
Green function in holography by employing the standard method in the
AdS/CFT dictionary. Since the fermion must be a gauge-invariant operator,
it could be a baryonic fermion. In particular, the influence of the
instanton charge denoted by $q$ in the fermionic correlation function
is also considered.

\subsection{General setup for the fermionic correlation function}

Let us start with the principle of the AdS/CFT which postulates that
the generating functional of the dual conformal field theory (CFT)
$Z_{\mathrm{CFT}}$ is equal to its gravitational partition function
$Z_{\mathrm{gravity}}$ in the bulk geometry. Namely, for a spinor
field $\psi$ in the $D+1$ dimensional bulk, we have\cite{key-b2,key-b7,key-b1}

\begin{equation}
Z_{\mathrm{CFT}}\left[\bar{\psi}_{0},\psi_{0}\right]=Z_{\mathrm{gravity}}\left[\bar{\psi},\psi\right]\big|_{\bar{\psi},\psi\rightarrow\bar{\psi}_{0},\psi_{0}},\label{eq:27}
\end{equation}
with

\begin{align}
Z_{\mathrm{CFT}}\left[\bar{\psi_{0}},\psi_{0}\right] & =\left\langle \exp\left\{ \int_{\partial\mathcal{M}}\left(\bar{\eta}\psi_{0}+\bar{\psi}_{0}\eta\right)d^{D}x\right\} \right\rangle ,\nonumber \\
Z_{\mathrm{gravity}}\left[\bar{\psi},\psi\right] & =e^{-S_{\mathrm{gravity}}^{ren}},\nonumber \\
S_{\mathrm{gravity}}^{ren} & =\int_{\mathcal{M}}\mathcal{L}_{\mathrm{gravity}}^{ren}\left[\bar{\psi},\psi\right]\sqrt{-g}d^{D+1}x
\end{align}
Note that, we use $\mathcal{M}$ to denote the bulk space and its
holographic boundary is given at $\partial\mathcal{M}=\left\{ z\rightarrow0\right\} $.
$\psi_{0}$ refers to the boundary value of $\psi$ as a source of
the boundary fermionic operator $\eta$. Since $\psi,\eta$ are fermionic
gauge-invariant operator, $\eta$ is expected to be a baryon or a
baryonic plasmino phenomenologically in this work. $\mathcal{L}_{\mathrm{gravity}}^{ren}$
denotes the renormalized Lagrangian of the bulk field $\psi$. Then
following the standard steps in quantum field theory (QFT), the one-point
function i.e. average value of $\bar{\eta}$ is given as

\begin{equation}
\left\langle \bar{\eta}\right\rangle =\frac{1}{Z_{\mathrm{CFT}}}\frac{\delta Z_{\mathrm{CFT}}}{\delta\psi_{0}}=\frac{1}{Z_{\mathrm{gravity}}}\frac{\delta Z_{\mathrm{gravity}}}{\delta\psi_{0}}=-\frac{\delta S_{\mathrm{gravity}}^{ren}}{\delta\psi_{0}}\equiv\Pi_{0},\label{eq:29-2}
\end{equation}
where we have used the relation (\ref{eq:27}). Therefore, the two-point
correlation function $G_{R}$ of $\eta$ is obtained as,

\begin{equation}
\left\langle \bar{\eta}\left(\omega,\vec{k}\right)\right\rangle =G_{R}\left(\omega,\vec{k}\right)\psi_{0},
\end{equation}
i.e.

\begin{equation}
\Pi_{0}=G_{R}\left(\omega,\vec{k}\right)\psi_{0},\label{eq:31}
\end{equation}
where $\omega,\vec{k}$ refers to the frequency and 3-momentum of
the associated Fourier modes. Altogether, it is possible to evaluate
$\psi_{0},\Pi_{0}$ in order to evaluate the two-point correlation
function $G_{R}$ of $\eta$ by using the classical gravity action
$S_{\mathrm{gravity}}^{ren}$ in holography.

\subsection{Covariant Dirac equation for a probe fermion}

Keep the formulas in Section 4.1 in hand, let us investigate a probe
fermion in the D+1-dimensional holographic background (\ref{eq:14}).
In general, the D+1-dimensional homogeneous background metric can
be written as,

\begin{equation}
ds_{\left(5\mathrm{d}\right)}^{2}=g_{tt}dt^{2}+g_{xx}d\mathbf{x}\cdot d\mathbf{x}+g_{zz}dz^{2},\label{eq:32}
\end{equation}
where $g_{tt},g_{xx},g_{zz}$ depend on $z$ only. For a probe fermion
in bulk, its dynamic is described by the Dirac action given in (\ref{eq:17}),
so the Dirac equation reads

\begin{equation}
\left(\Gamma^{M}\nabla_{M}-m\right)\psi=0,\label{eq:33}
\end{equation}
where the covariant derivative operator is given in terms of the spin
connection as,

\begin{equation}
\nabla_{M}=\partial_{M}+\frac{1}{4}\omega_{Mab}\gamma^{ab},\gamma^{ab}=\frac{1}{2}\left[\gamma^{a},\gamma^{b}\right].
\end{equation}
Imposing the homogeneous metric (\ref{eq:32}), the Dirac operator
is computed as,

\begin{equation}
\Gamma^{M}\nabla_{M}=\frac{1}{\sqrt{-g_{tt}}}\gamma^{0}\partial_{0}+\frac{1}{\sqrt{g_{xx}}}\gamma^{i}\partial_{i}+\frac{1}{\sqrt{g_{zz}}}\gamma\partial_{z}+\frac{1}{4g_{xx}g_{tt}\sqrt{g_{zz}}}\left[\left(D-1\right)g_{tt}g_{xx}^{\prime}+g_{xx}g_{tt}^{\prime}\right]\gamma.
\end{equation}
To find a solution for the Dirac equation, we insert the ansatz for
spinor $\psi$ in Dirac representation as,

\begin{align}
\psi & =\left(\begin{array}{c}
\psi_{R}\\
\psi_{L}
\end{array}\right)=\left(-gg^{zz}\right)^{-1/4}\int\frac{d^{4}p}{\left(2\pi\right)^{4}}e^{ik\cdot x}\chi\left(z,k\right),\nonumber \\
\chi\left(z,k\right) & =\left[\begin{array}{c}
\chi_{R}\left(z,k\right)\\
\chi_{L}\left(z,k\right)
\end{array}\right],\chi_{R,L}=\left(\begin{array}{c}
\chi_{R,L}^{\left(1\right)}\\
\chi_{R,L}^{\left(2\right)}
\end{array}\right),k_{\mu}=\left(k_{0},k_{i}\right),\label{eq:36}
\end{align}
into the Dirac equation (\ref{eq:33}), so it is simplified as,

\begin{equation}
\left[\sqrt{\frac{g_{xx}}{g_{zz}}}\left(\gamma\partial_{z}-m\sqrt{g_{zz}}\right)+iK_{\mu}\gamma^{\mu}\right]\chi\left(z,k\right)=0,\label{eq:37}
\end{equation}
where

\begin{equation}
K_{\mu}=\left(\sqrt{-\frac{g_{xx}}{g_{tt}}}k_{0},k_{i}\right).
\end{equation}

\subsection{Construction for the correlation function}

When the background metric (\ref{eq:32}) is inserted into (\ref{eq:37}),
the Dirac equation reduces to two decoupled differential equations
for $\chi_{R,L}$ as,

\begin{align}
\left(\partial_{z}+m\sqrt{g_{zz}}\right)\left[\left(K\cdot\sigma\right)^{-1}\sqrt{\frac{g_{xx}}{g_{zz}}}\left(\partial_{z}-m\sqrt{g_{zz}}\right)\chi_{R}\right] & =-\sqrt{\frac{g_{zz}}{g_{xx}}}\left(K\cdot\bar{\sigma}\right)\chi_{R},\nonumber \\
\left(\partial_{z}-m\sqrt{g_{zz}}\right)\left[\left(K\cdot\bar{\sigma}\right)^{-1}\sqrt{\frac{g_{xx}}{g_{zz}}}\left(\partial_{z}+m\sqrt{g_{zz}}\right)\chi_{L}\right] & =-\sqrt{\frac{g_{zz}}{g_{xx}}}\left(K\cdot\sigma\right)\chi_{L},
\end{align}
which can be solved analytically near the boundary $z\rightarrow0$
as,

\begin{align}
\chi_{R} & =Cz^{1-mR}+Dz^{mR},\nonumber \\
\chi_{L} & =Az^{-mR}+Bz^{1+mR}.\label{eq:40}
\end{align}
where $A,B,C,D$ are constant Weyl spinor depending on 4-momentum
$k$ and the charge of the D-instanton $q$. Hence for $mR>0$, the
boundary value $\chi_{0}$ of $\chi$ is given by $A$ which is the
most divergent term, defined as

\begin{equation}
\chi_{0}=\lim_{z\rightarrow0}z^{mR}\chi=\left(\begin{array}{c}
0\\
A
\end{array}\right).\label{eq:41}
\end{equation}
Then recall the Dirac action (\ref{eq:17}) with (\ref{eq:36}), it
leads to

\begin{align}
S_{\mathrm{bulk}} & =i\int d^{4}xdz\sqrt{-g}\bar{\psi}\left(\Gamma^{M}\nabla_{M}-m\right)\psi\nonumber \\
 & =i\int d^{4}x\left(\bar{\chi}\gamma\chi\right)|_{z_{H}}^{0}-i\int d^{4}xdz\sqrt{\frac{g_{zz}}{g_{xx}}}\left[\sqrt{\frac{g_{xx}}{g_{zz}}}\left(\partial_{z}\bar{\chi}\gamma+m\sqrt{g_{zz}}\bar{\chi}\right)+iK_{\mu}\bar{\chi}\gamma^{\mu}\right]\chi.\label{eq:42}
\end{align}
Note that the last term in (\ref{eq:42}) vanishes since it is nothing
but the conjugate equation of (\ref{eq:37}). Therefore, we can obtain
a boundary action $S_{\mathrm{bdry}}$ from (\ref{eq:42}) as,

\begin{align}
S_{\mathrm{bdry}} & =i\int d^{4}x\left(\bar{\chi}\gamma\chi\right)|_{z\rightarrow0}\nonumber \\
 & =i\int d^{4}x\left(D^{\dagger}A+C^{\dagger}Az^{1-2m}\right)|_{z\rightarrow0},
\end{align}
which implies the holographic counterterm $S_{\mathrm{CT}}$ at the
boundary is,

\begin{equation}
S_{\mathrm{CT}}=-i\int d^{4}xC^{\dagger}Az^{1-2m}|_{z\rightarrow0},
\end{equation}
giving the holographic renormalized action as,

\begin{align}
S_{\mathrm{gravity}}^{ren} & =S_{\mathrm{bdry}}+S_{\mathrm{CT}}\nonumber \\
 & =i\int d^{4}xD^{\dagger}A.\label{eq:45}
\end{align}
Picking up (\ref{eq:45}) with (\ref{eq:29-2}) and (\ref{eq:31}),
the correlation function satisfies

\begin{equation}
D=-G_{R}A,\label{eq:46}
\end{equation}
in the Weyl representation. This is not surprising since the boundary
value $\psi_{0}$ of $\psi$ is in fact a chiral spinor. On the other
hand, the correlation function in the $2\times2$ matrix must be able
to be written as the combination of the Pauli matrices as,

\begin{equation}
G_{R}=\alpha+\beta k_{i}\sigma^{i},
\end{equation}
we can set that the momentum is along the $x^{3}$ direction as $k_{\mu}=\left(-\omega,0,0,\mathrm{k}\right)$
due to the rotational symmetry in $\mathbb{R}^{3}=\left\{ x^{i}\right\} $,
so that the correlation function $G_{R}$ is expected to be diagonal,
as

\begin{equation}
G_{R}=\left(\begin{array}{cc}
G_{R}^{1,1} & 0\\
0 & G_{R}^{2,2}
\end{array}\right).
\end{equation}

\begin{figure}[t]
\begin{centering}
\includegraphics[scale=0.5]{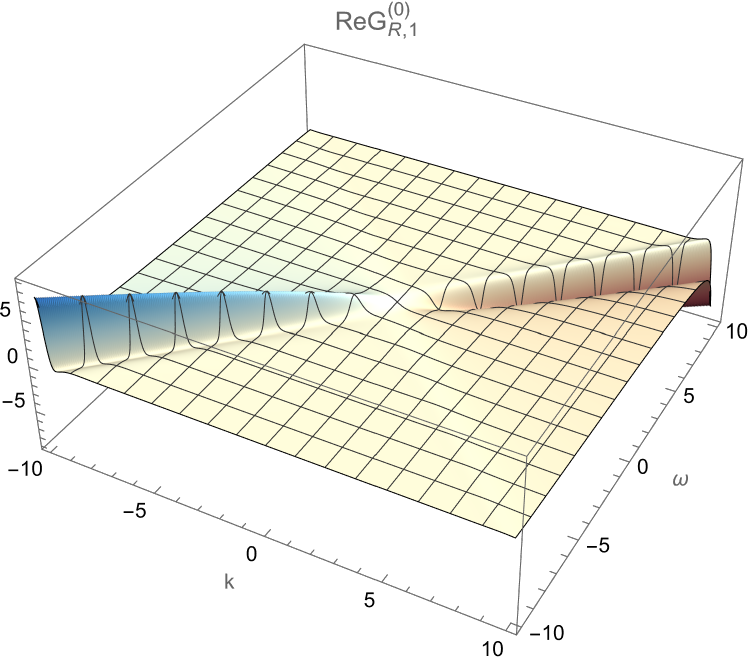}\includegraphics[scale=0.5]{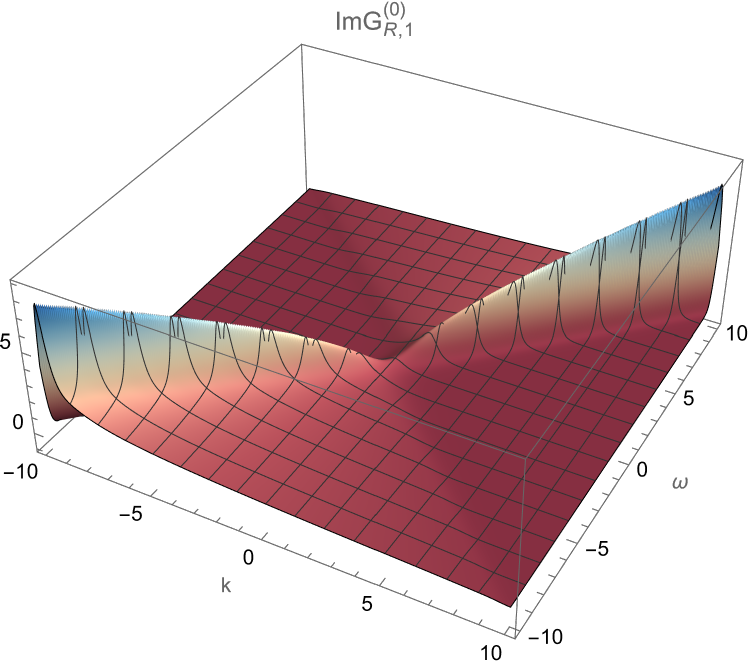}
\par\end{centering}
\begin{centering}
\includegraphics[scale=0.51]{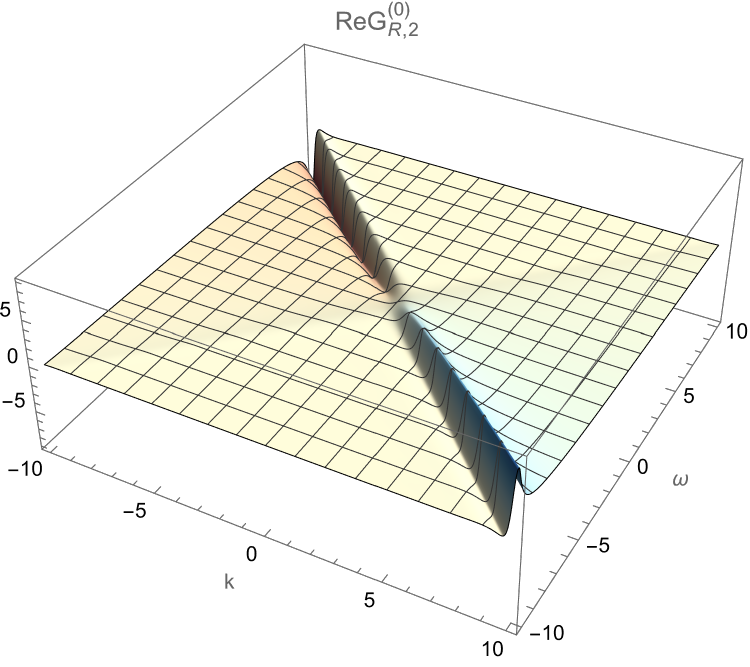}\includegraphics[scale=0.51]{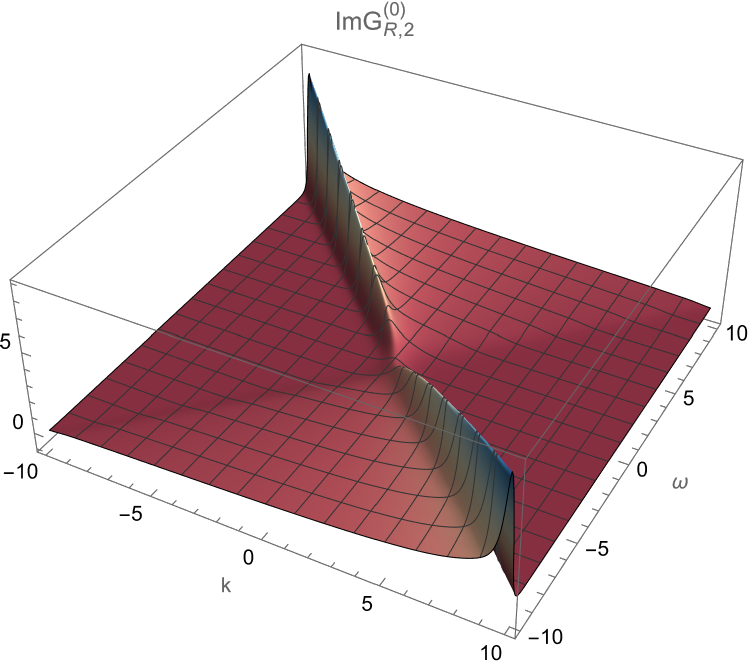}
\par\end{centering}
\caption{\label{fig:2}The real and imaginary parts of zero-th order correlation
functions $G_{R,\alpha}^{\left(0\right)}$ as functions of $\omega,\mathrm{k}$
in the deconfined phase. The parameter is chosen as $q=0,m=0.01,z_{H}=1.$}
\end{figure}
 
\begin{figure}[t]
\begin{centering}
\includegraphics[scale=0.5]{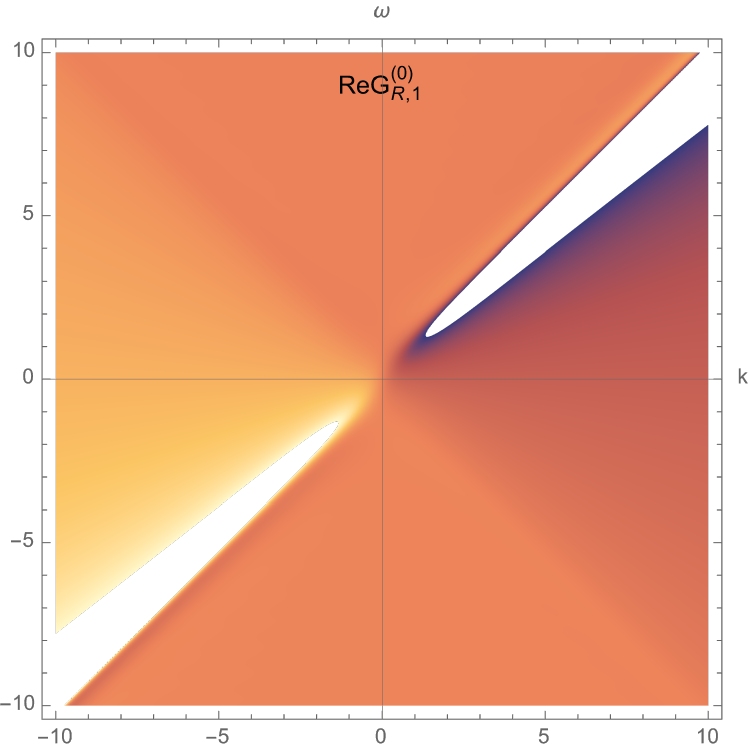}\includegraphics[scale=0.5]{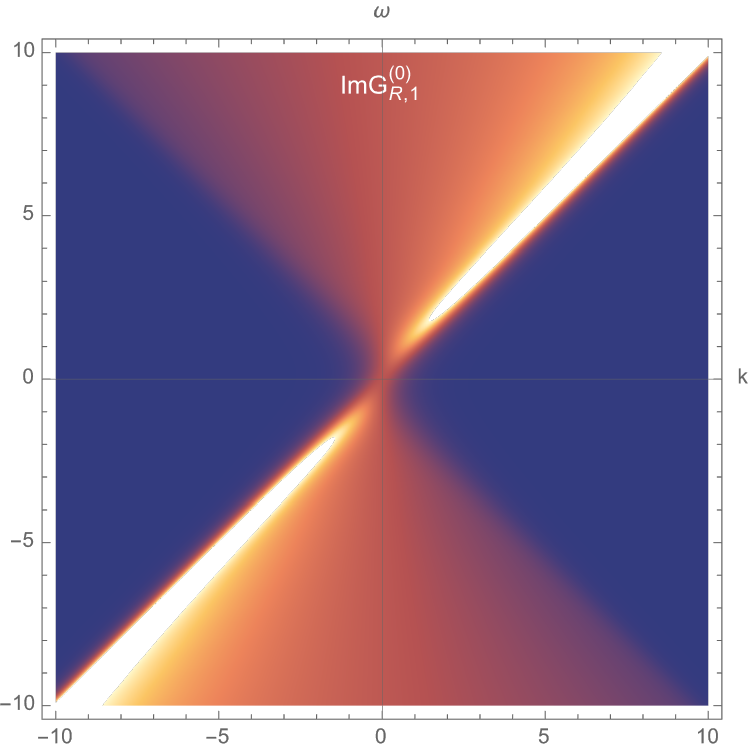}
\par\end{centering}
\begin{centering}
\includegraphics[scale=0.5]{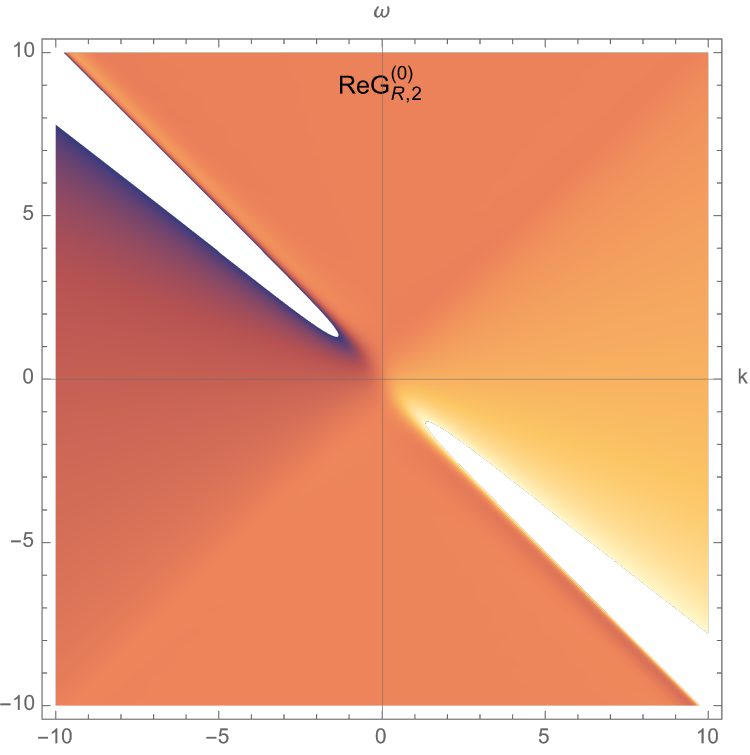}\includegraphics[scale=0.5]{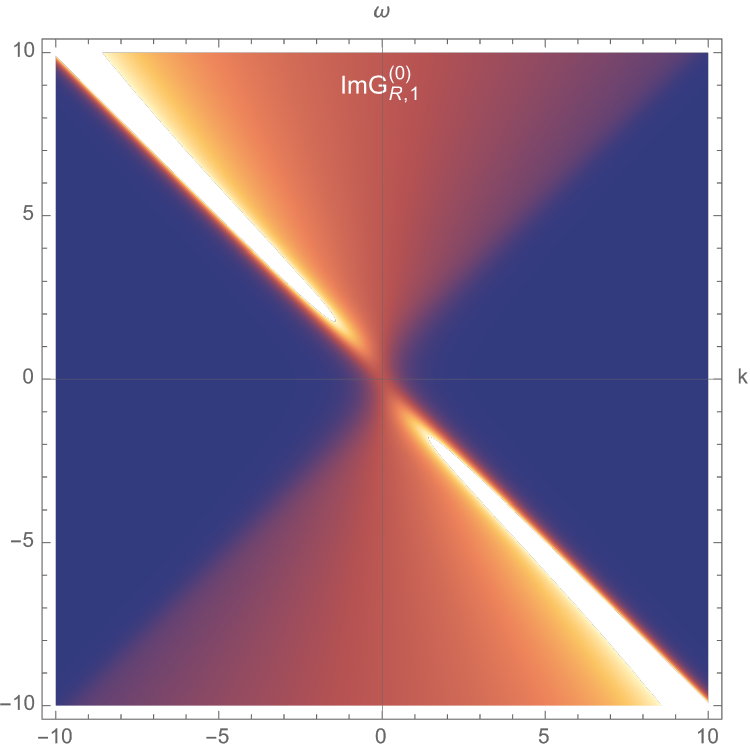}
\par\end{centering}
\caption{\label{fig:3}Density plots of the zero-th order correlation functions
$G_{R,\alpha}^{\left(0\right)}$ in the deconfined phase with the
parameters as $q=0,m=0.01,z_{H}=1.$ The white regions refer to the
peaks in the correlation functions.}
\end{figure}
To finalize this section, we introduce the ratio defined as

\begin{equation}
G_{\alpha}=\left(-1\right)^{\alpha+1}\frac{\chi_{R}^{\left(\alpha\right)}}{\chi_{L}^{\left(\alpha\right)}},\ \alpha=1,2.
\end{equation}
According to (\ref{eq:41}) - (\ref{eq:46}), the correlation function
is given by

\begin{equation}
G_{R}^{\alpha,\alpha}=\left(-1\right)^{\alpha}\lim_{z\rightarrow0}z^{-2m}G_{\alpha},\label{eq:50}
\end{equation}
and the equations for the ratios can be obtained by recombining the
components of the spinor $\chi$ with respect to the Dirac equation
(\ref{eq:37}). As a result, one can find the equations for the ratios
are given as,

\begin{equation}
\sqrt{\frac{g_{xx}}{g_{zz}}}G_{\alpha}^{\prime}=\left(-1\right)^{\alpha}u-\mathrm{k}+\left[\mathrm{k}+\left(-1\right)^{\alpha}u\right]G_{\alpha}^{2}+2m\sqrt{g_{xx}}G_{\alpha},\label{eq:51}
\end{equation}
where ``$^{\prime}$'' refers to the derivative with respect to
$z$, and

\begin{equation}
u=\omega\sqrt{-\frac{g_{xx}}{g_{tt}}}=\frac{\omega}{\sqrt{f}}.
\end{equation}
Therefore the correlation function could be evaluated numerically
by solving (\ref{eq:51}) with the infalling boundary condition $G_{\alpha}|_{z=z_{H}}=\left(-1\right)^{\alpha}i$
obtained by analyzing its asymptotic behavior near $z=z_{H}$ as \cite{key-35,key-36,key-37,key-38,key-39}. 

\subsection{The numerical analysis}

\subsubsection*{Deconfined phase}

\begin{figure}[t]
\begin{centering}
\includegraphics[scale=0.52]{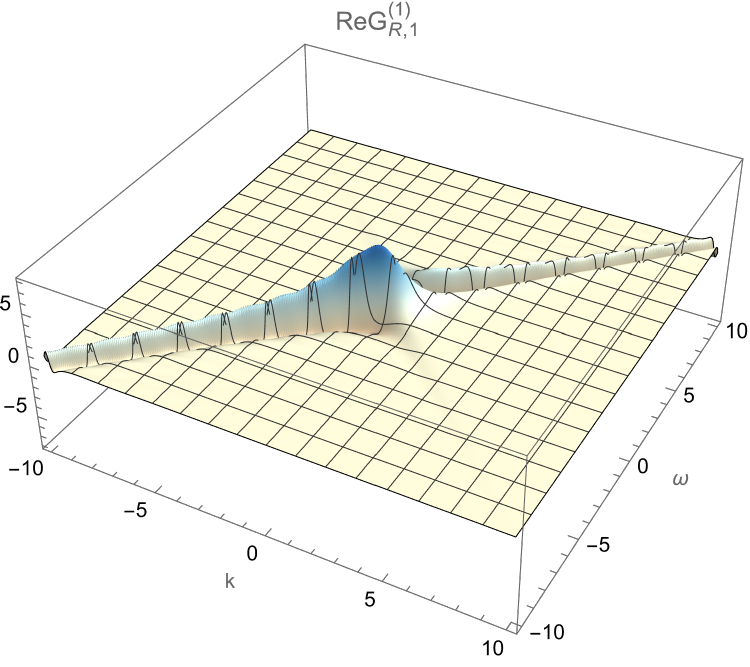}\includegraphics[scale=0.5]{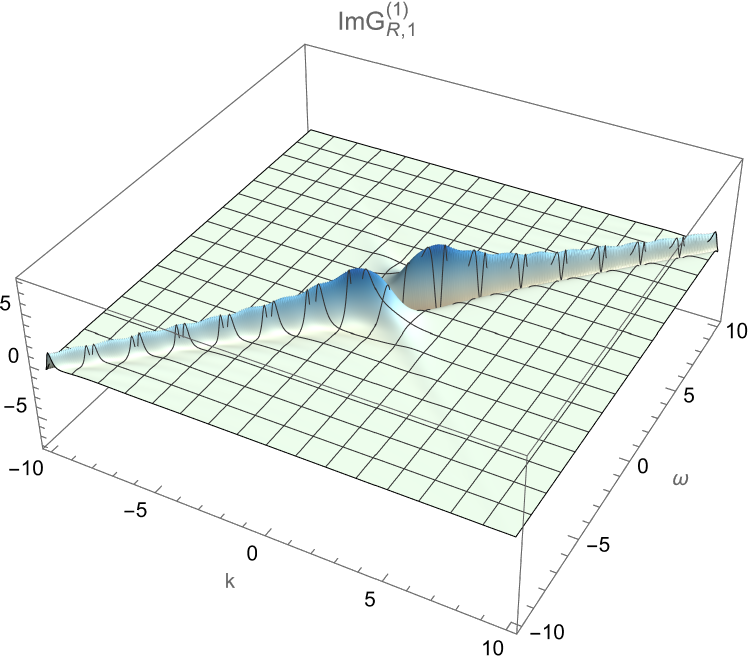}
\par\end{centering}
\begin{centering}
\includegraphics[scale=0.5]{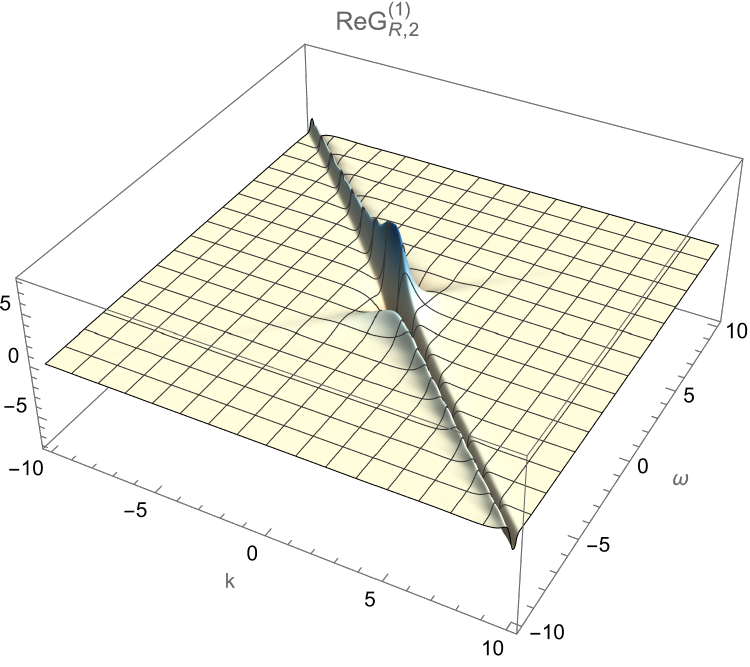}\includegraphics[scale=0.5]{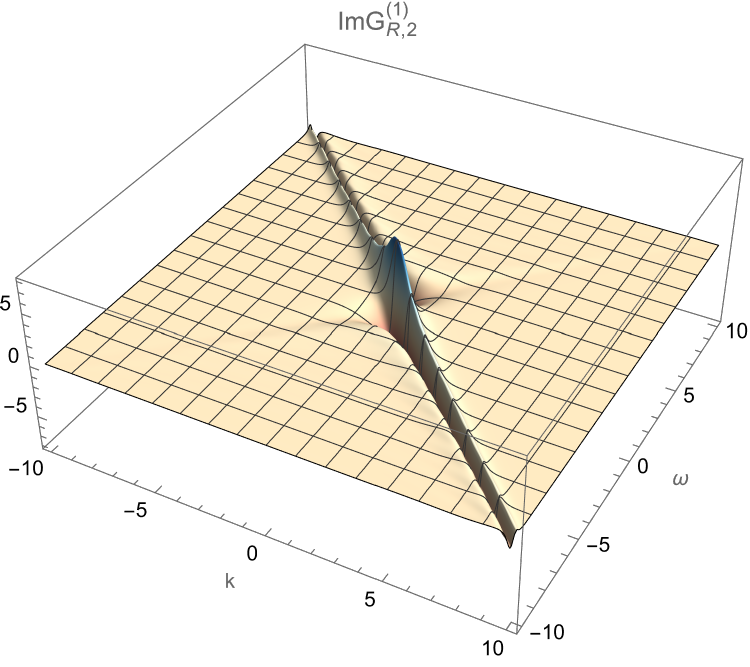}
\par\end{centering}
\caption{\label{fig:4}The real and imaginary parts of leading order deconfined
correlation functions $G_{R,\alpha}^{\left(1\right)}$ as functions
of $\omega,\mathrm{k}$ with $m=0.01,z_{H}=1.$}
\end{figure}

\begin{figure}[t]
\begin{centering}
\includegraphics[scale=0.5]{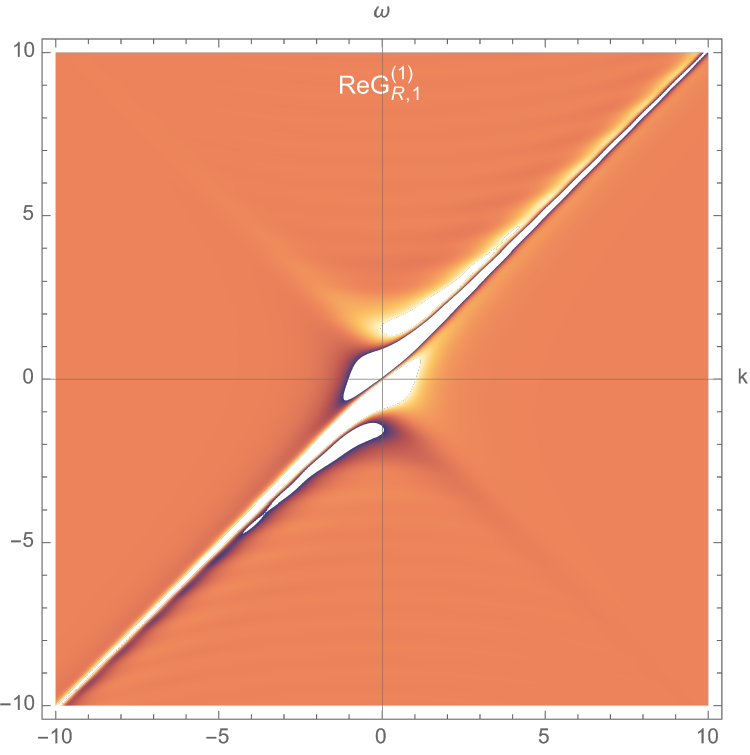}\includegraphics[scale=0.5]{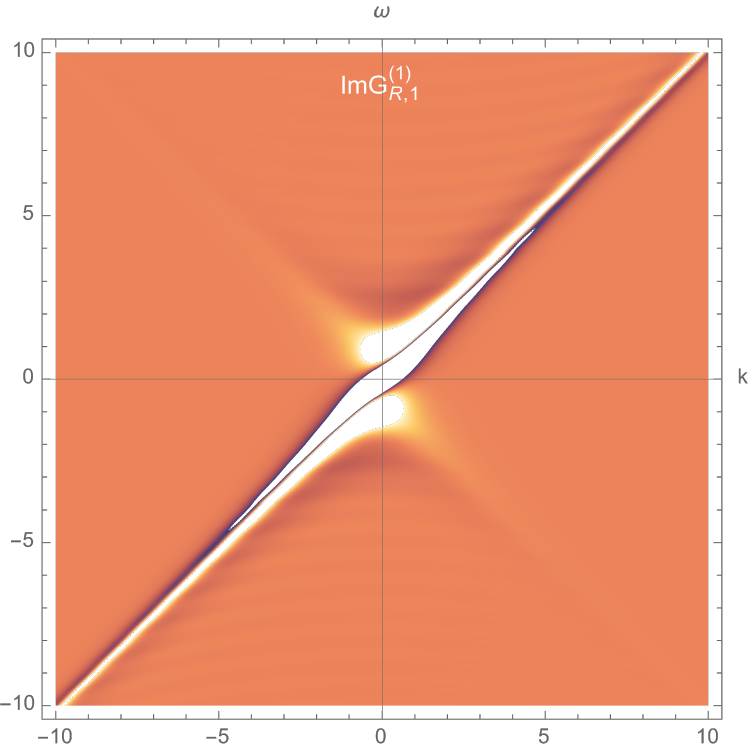}
\par\end{centering}
\begin{centering}
\includegraphics[scale=0.5]{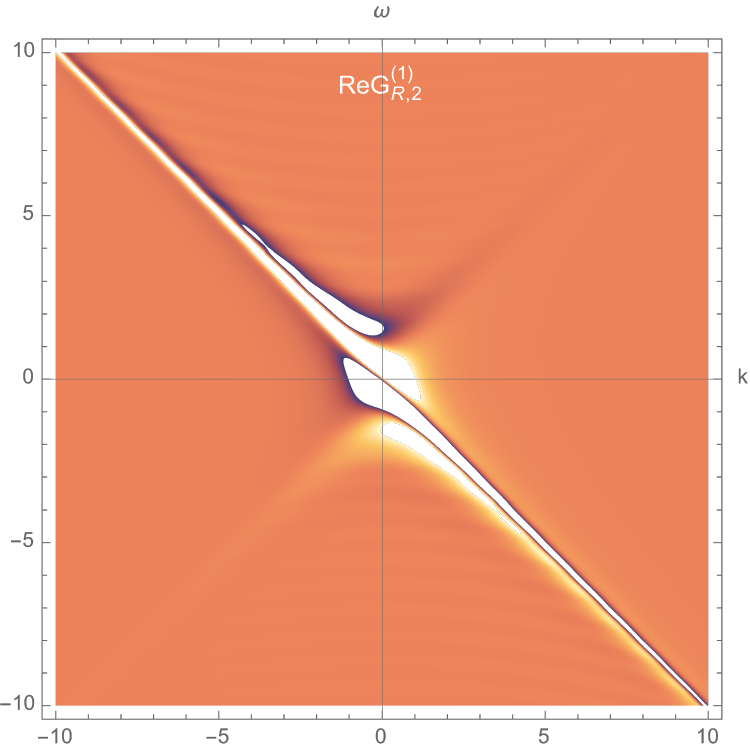}\includegraphics[scale=0.5]{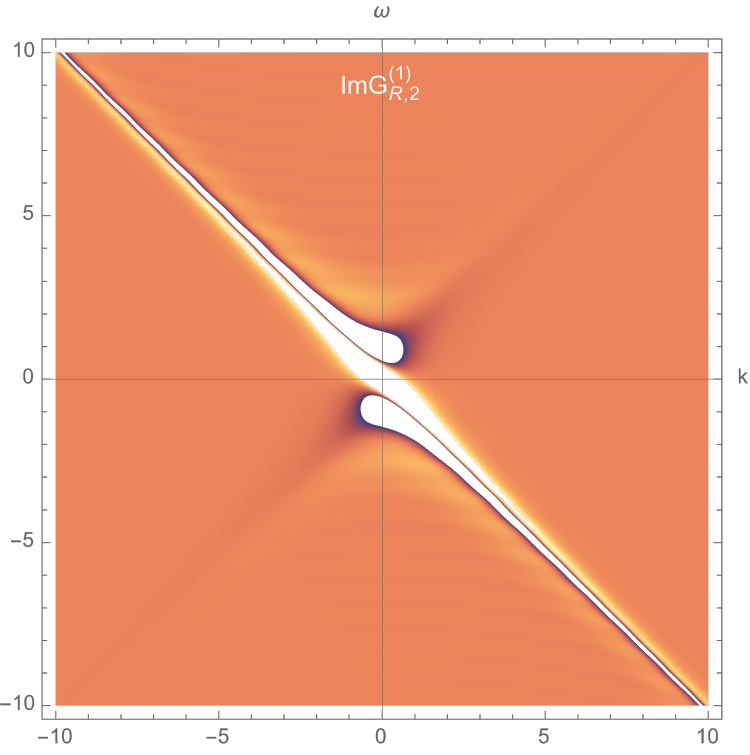}
\par\end{centering}
\caption{\label{fig:5}Density plots of the leading order deconfined correlation
functions $G_{R,\alpha}^{\left(1\right)}$ as functions of $\omega,\mathrm{k}$
with $m=0.01,z_{H}=1.$ The white regions refer to the peaks in the
correlation functions}
\end{figure}

\begin{figure}[t]
\begin{centering}
\includegraphics[scale=0.36]{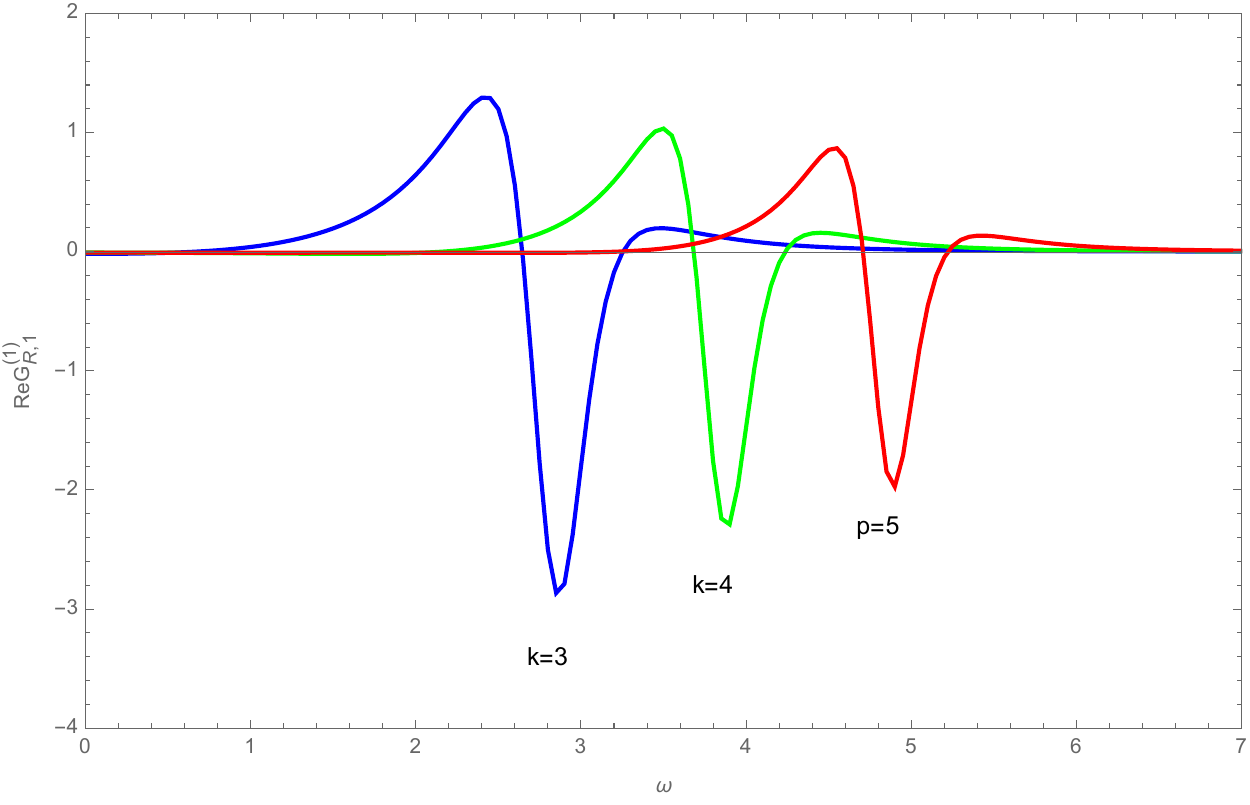}\includegraphics[scale=0.36]{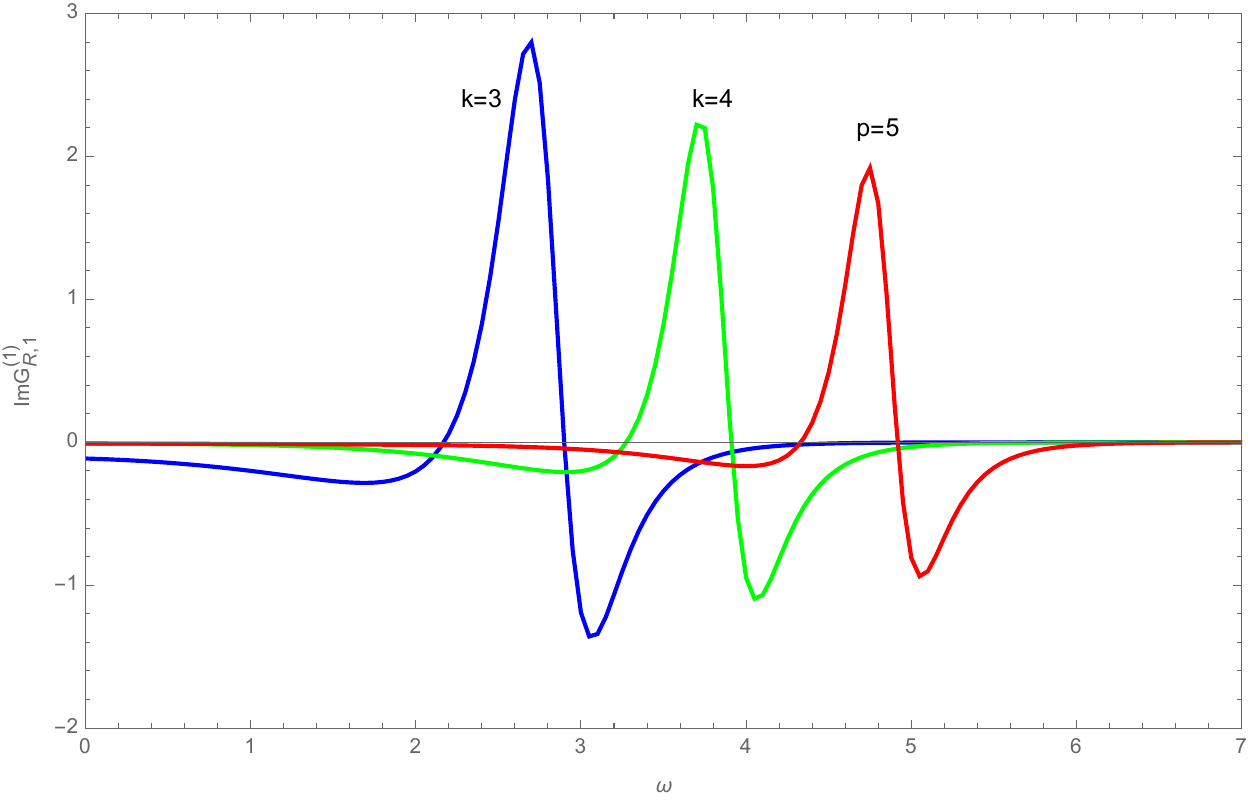}
\par\end{centering}
\begin{centering}
\includegraphics[scale=0.36]{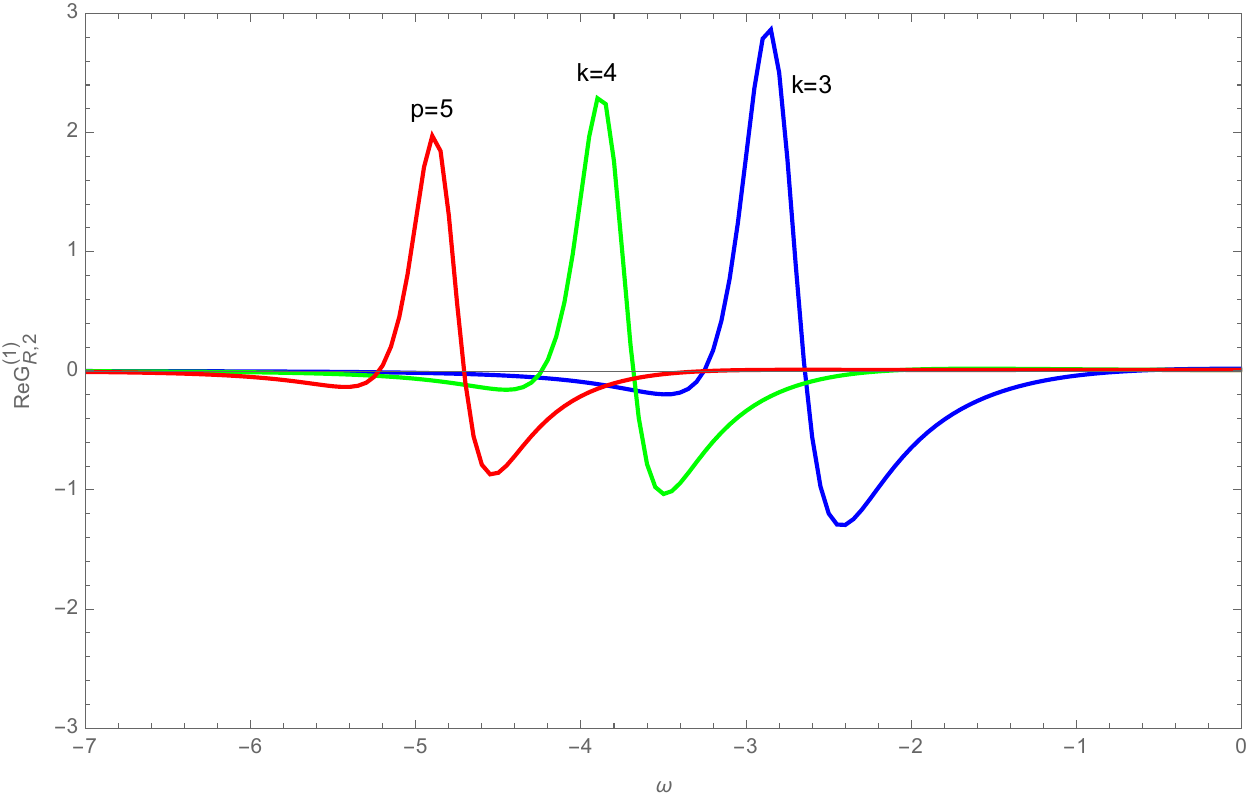}\includegraphics[scale=0.36]{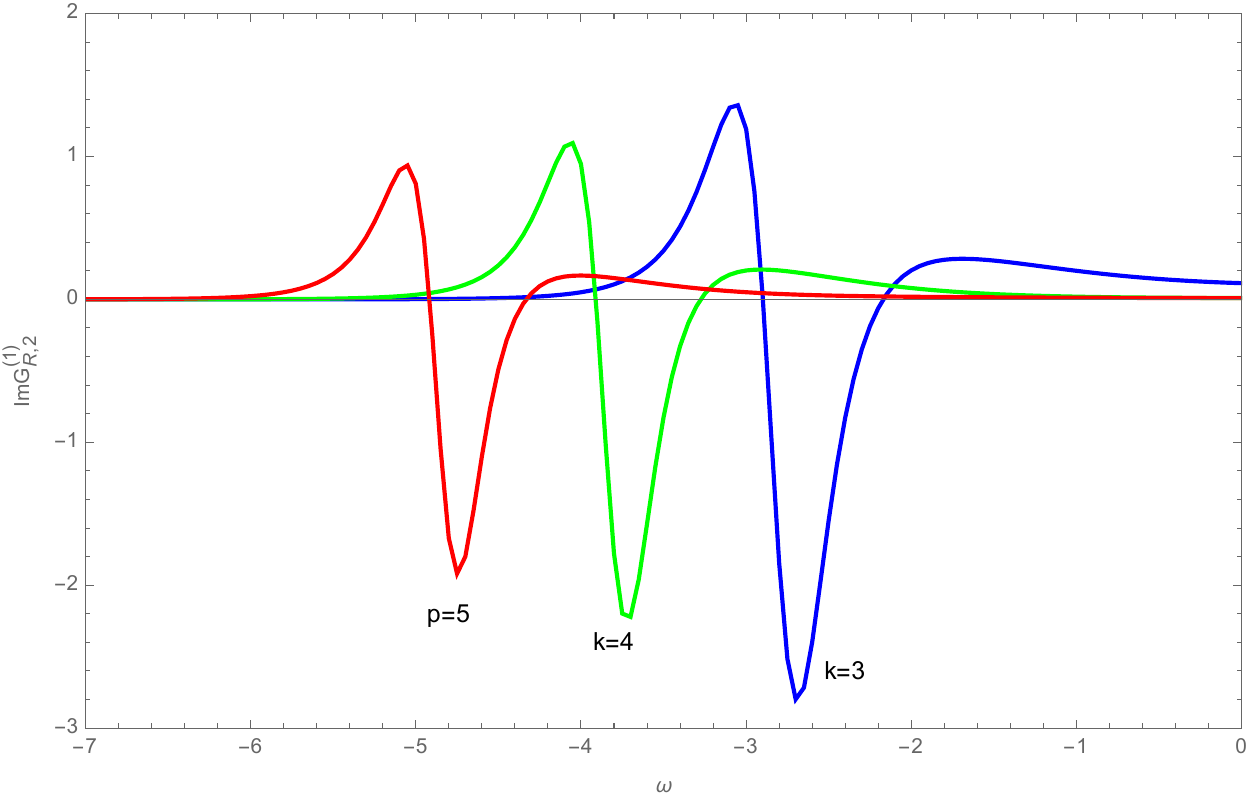}
\par\end{centering}
\caption{\label{fig:6}The real and imaginary parts of leading order deconfined
correlation functions $G_{R,\alpha}^{\left(1\right)}$ as functions
of $\omega$ with various $\mathrm{k}$ and $m=0.01,z_{H}=1$. There
are always two peaks for various $\mathrm{k}$ in the correlation
function which correspond to the two branches of the dispersion curves.}
\end{figure}

When the D-instanton is taken into account, the correlation function
given in (\ref{eq:50}) should also depend on the charge density of
the D-instanton denoted by $q$. In order to evaluate exactly the
$q$-dependence in the correlation function, for $q\ll1$ we consider
its leading order correction by expand $G_{\alpha}$ as,

\begin{equation}
G_{\alpha}=\mathcal{G}_{\alpha}+q\xi_{\alpha},
\end{equation}
where $\mathcal{G}_{\alpha}$ is the zero-th order solution satisfying
the equations presented in (\ref{eq:51}) with $q=0$ as (there is
no sum for the repetitive indices),

\begin{equation}
\sqrt{f}\mathcal{G}_{\alpha}^{\prime}=\left(-1\right)^{\alpha}u-\mathrm{k}+\left[\mathrm{k}+\left(-1\right)^{\alpha}u\right]\mathcal{G}_{\alpha}^{2}+\frac{2mR}{z}\mathcal{G}_{\alpha}.\label{eq:54}
\end{equation}
Here we have inserted the deconfined geometry (\ref{eq:14}) into
(\ref{eq:51}), and $\xi_{\alpha}$ is the leading order correction
satisfying the equations 

\begin{align}
\sqrt{f}\xi_{\alpha}^{\prime}= & \left[\mathrm{k}+\left(-1\right)^{\alpha}u\right]2\mathcal{G}_{\alpha}\xi_{\alpha}+\frac{2mR}{z}\left[\xi_{\alpha}-\frac{1}{4}\mathcal{G}_{\alpha}\ln f\right].\label{eq:55}
\end{align}
By analyzing the asymptotic behavior of (\ref{eq:54}), we find the
infalling boundary condition for $\mathcal{G}_{\alpha}$ could be
$\mathcal{G}_{\alpha}|_{z=z_{H}}=\left(-1\right)^{\alpha}i$ which
is exactly same as the infalling boundary condition for $G_{\alpha}$.
Therefore, the boundary condition for $\xi_{\alpha}$ is expected
to be $\xi_{\alpha}|_{z=z_{H}}=0$. Altogether, we will solve (\ref{eq:54})
and (\ref{eq:55}) numerically in order to evaluate the associated
correlation function $G_{R}$ as the fermionic spectral function in
holography. We write down the correlation functions presented in (\ref{eq:50})
as the combinations of its zero-th order $G_{R,\alpha}^{\left(0\right)}$
and first order solution $G_{R,\alpha}^{\left(1\right)}$ as,

\begin{align}
G_{R}^{\alpha,\alpha} & =G_{R,\alpha}^{\left(0\right)}+qG_{R,\alpha}^{\left(1\right)},\nonumber \\
G_{R,\alpha}^{\left(0\right)} & =\left(-1\right)^{\alpha}\lim_{z\rightarrow0}z^{-2m}\mathcal{G}_{\alpha},\nonumber \\
G_{R,\alpha}^{\left(1\right)} & =\left(-1\right)^{\alpha}\lim_{z\rightarrow0}z^{-2m}\xi_{\alpha},
\end{align}
where $\mathcal{G}_{\alpha},\xi_{\alpha}$ are coefficients independent
on $q$ and they can be solved numerically through the equations (\ref{eq:54})
and (\ref{eq:55}). The corresponding numerical results for $\mathcal{G}_{\alpha},\xi_{\alpha}$
are given in Figure \ref{fig:2} - Figure \ref{fig:6}. 
 
Let us compare our numerical results with the calculations by using
the method of hard thermal loop (HTL) in thermal QCD. The approach
of HTL with vanishing chemical potential illustrates the dispersion
curves of fermions (plasmino) is \footnote{We pick up the modes of positive frequency with $\omega>0$ and momentum
$\mathrm{k}>0$, the case for the negative frequency can be obtained
by replacing $\omega\rightarrow-\omega$.}\cite{key-41},

\begin{align}
\omega\left(\mathrm{k}\right) & \simeq m_{f}\pm\frac{1}{3}\mathrm{k}+\frac{1}{3m_{f}}\mathrm{k}^{2},\ \mathrm{k}\ll1;\nonumber \\
\omega\left(\mathrm{k}\right) & \simeq\mathrm{k},\ \mathrm{k}\gg1,\label{eq:57}
\end{align}
where $m_{f}$ is the effective mass generated by the medium effect
of fermion as,

\begin{equation}
m_{f}=\sqrt{\frac{C_{F}}{8}}g_{\mathrm{YM}}T.
\end{equation}
Here $g_{\mathrm{YM}}$ refers to the Yang-Mills coupling constant
and $C_{F}$ is a constant suggested to be $C_{F}=4/3$ for fundamental
quarks or $C_{F}=1$ for the electron. Accordingly, we can see the
dispersion curves presented in the zero-th order correlation function
given in Figure \ref{fig:2} and Figure \ref{fig:3} cover basically
the behavior of dispersion curves from HTL at large momentum. For
the small momentum, while the zero-th order dispersion curves may
be close to the relation given in (\ref{fig:6}), it does not illustrate
two branches of the dispersion curves as some related works in holography
\cite{key-35,key-36,key-37,key-38,key-39}. The reason may be that,
according to the gravity/fluid correspondence \cite{key-42,key-43},
the plasma described holographically by the metric (\ref{eq:13})
with $q=0$ corresponds to the conformally ideal fluid of $\mathcal{N}=4$
super Yang-Mills theory which is different from the QED or QCD plasma.
In particular, the interaction of the plasmino or the dissipation
of the medium is not taken into account in the ideal fluid. While
this issue may be expected to be figured out by taking into account
the dissipation of the fluid somehow in the holographic background
(\ref{eq:13}), we will focus on the influence on the instanton in
this work.

For the case $q>0$, the correlation function has a correction term
$G_{R,\alpha}^{\left(1\right)}$ given by $\xi_{\alpha}$ as it is
illustrated in Figure \ref{fig:4} - Figure \ref{fig:6}. For $\omega>0$
and $\mathrm{k}>0$, it is clear to see that the dispersion curves
have two branches at small momentum and they trend to merge for large
momentum. By taking a look at the value of the correction terms, we
can find the correction $G_{R,\alpha}^{\left(1\right)}$ is large/small
for small/large momentum respectively which has an opposite behavior
to the zero-th correlation function $G_{R,\alpha}^{\left(0\right)}$.
Altogether, for $q>0$, the first order correction $G_{R,\alpha}^{\left(1\right)}$
dominates the behavior of the dispersion curves at small momentum
and it always manifests two branches. At large momentum, the zero-th
order function $G_{R,\alpha}^{\left(0\right)}$ dominates the behavior
of the dispersion curves which displays only one branch of the dispersion
curves on the light-cone. Therefore, the dispersion curves given by
the total spectral function $G_{R}^{\alpha,\alpha}$ are very close
to the relation for HTL (\ref{eq:57}). 

Besides, our numerical results also illustrate that, i.e. for $\omega,\mathrm{k}>0$,
the effective mass $m_{f}=\omega\left(\mathrm{k}=0\right)$ splits
into two values as it is displayed in Figure \ref{fig:5}. This may
be related remarkably to the properties of the fermionic spin, since
it is known that there is spin-dependent interactions among fermions
induced by instantons in QCD \cite{key-1,key-2}. According to the
calculations by one-gluon exchange, the massive quark potential depends
on the quantum number of spin in the presence of instanton, so the
total mass of the plasmino including quark interaction splits into
two values which leads to two branches in the dispersion. In this
sense, our numerical evaluation might be a holographic reproduction
of the instanton-induced interaction with spin. However, for the massless
case, the spin-dependent interaction induced by instantons trend to
vanish, so it could be why the spectral function of massless fermion
is not affected by the instanton charge. On the other hand, some existing
works about D-instanton in gauge-gravity duality \cite{key-24,key-28,key-29,key-30,key-31,key-32,key-33,key-35,key-36}
also support that the property of mass for a massless hadron in plasma
is not affected by the instanton, which could thus be a holographic
confirmation.

\subsubsection*{Confined phase}

\begin{figure}
\begin{centering}
\includegraphics[scale=0.22]{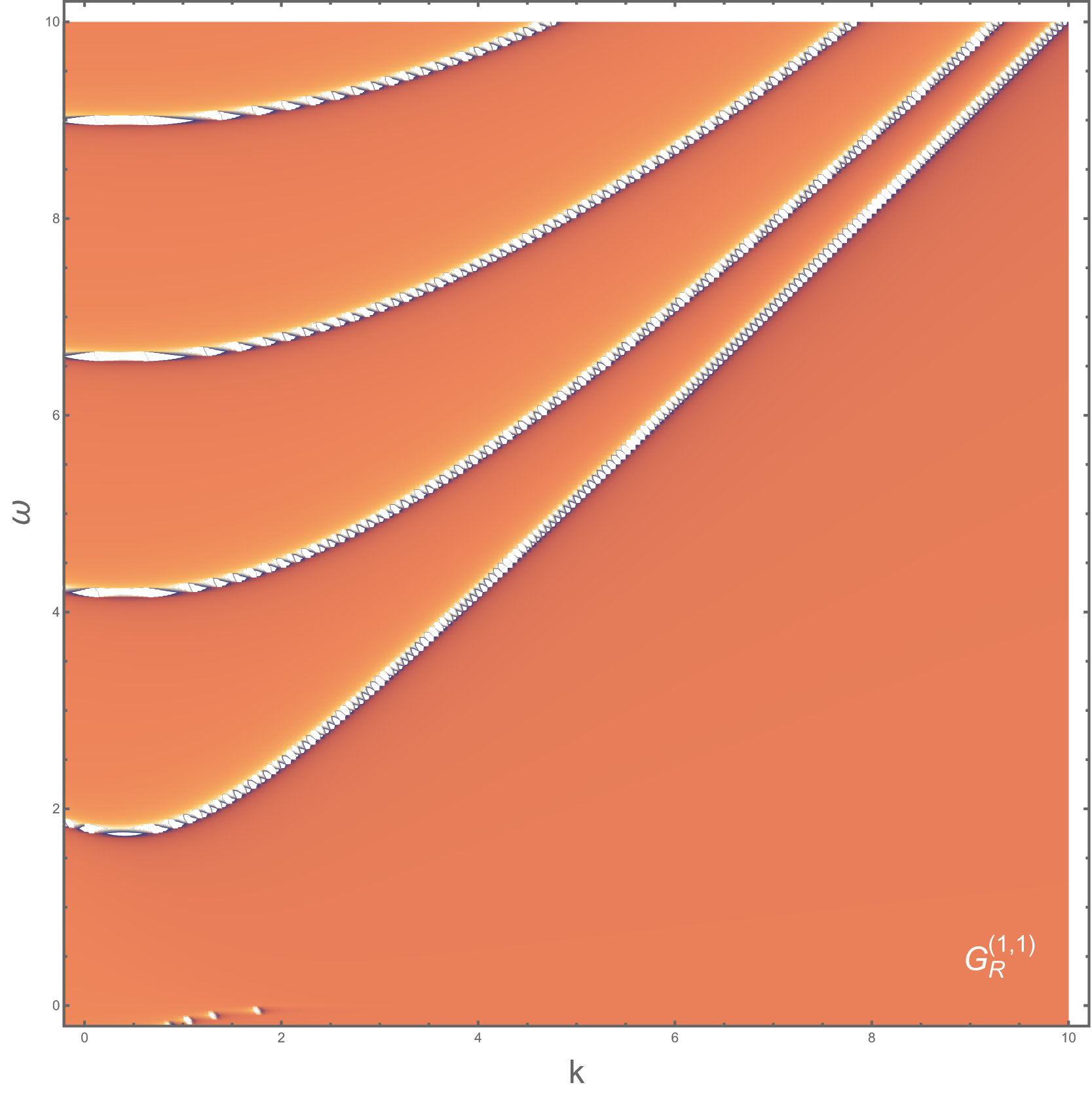}\includegraphics[scale=0.22]{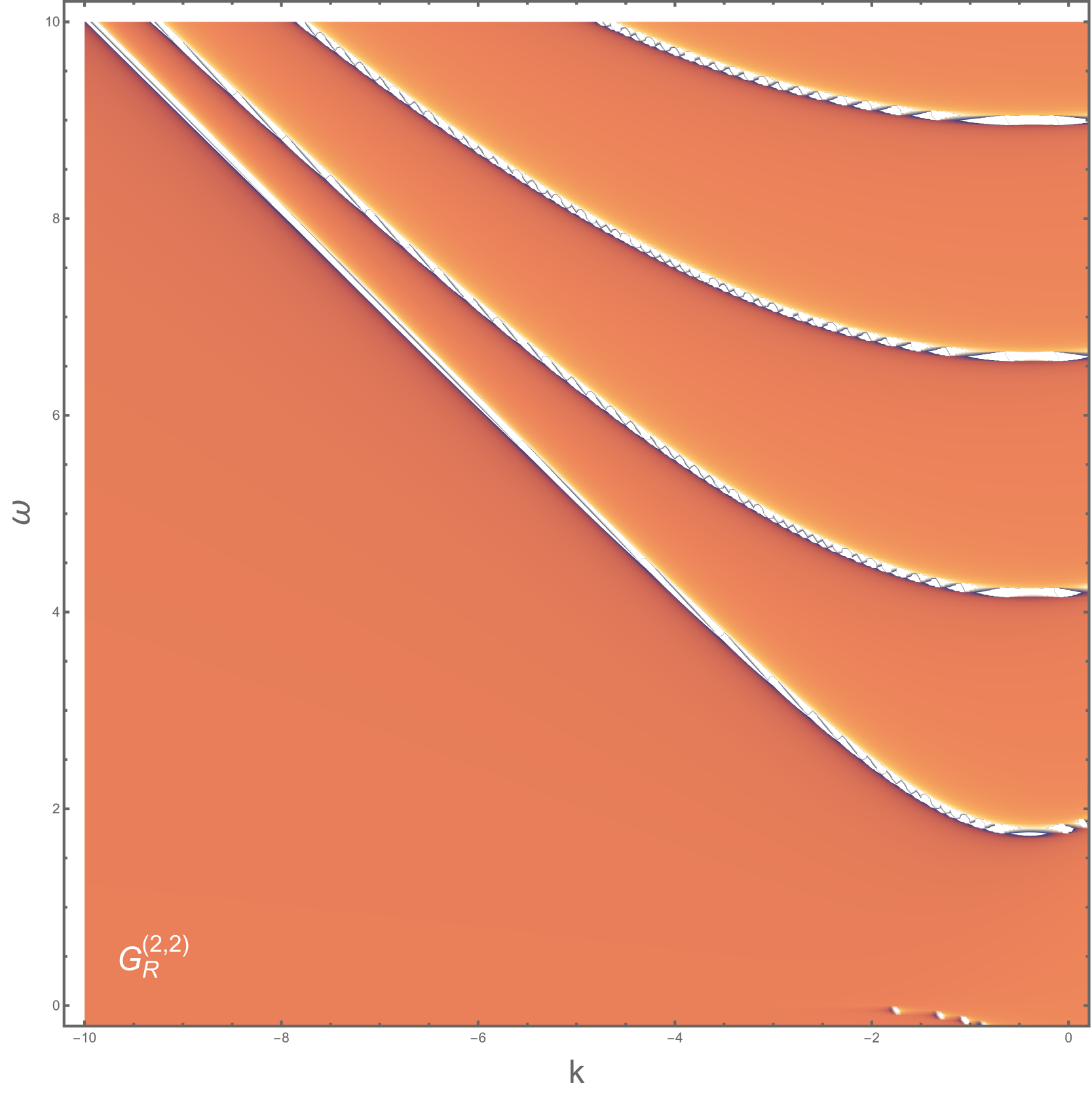}
\par\end{centering}
\caption{\label{fig:7}The density plot of the holographic correlation function
in confined phase with $q=1$. The white region refers to the peaks
in the correlation function.}
\end{figure}
 
\begin{figure}
\begin{centering}
\includegraphics[scale=0.27]{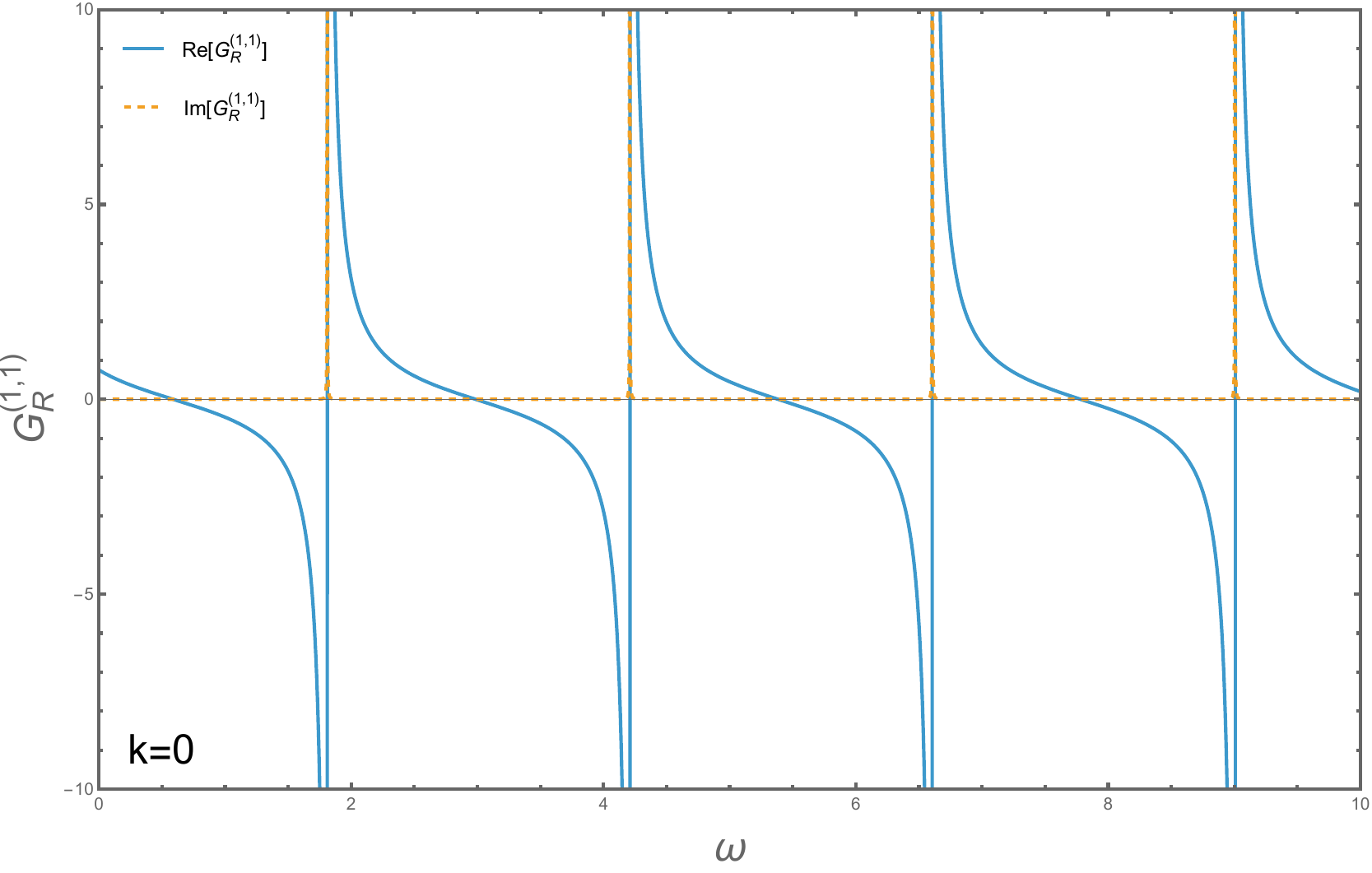}\includegraphics[scale=0.27]{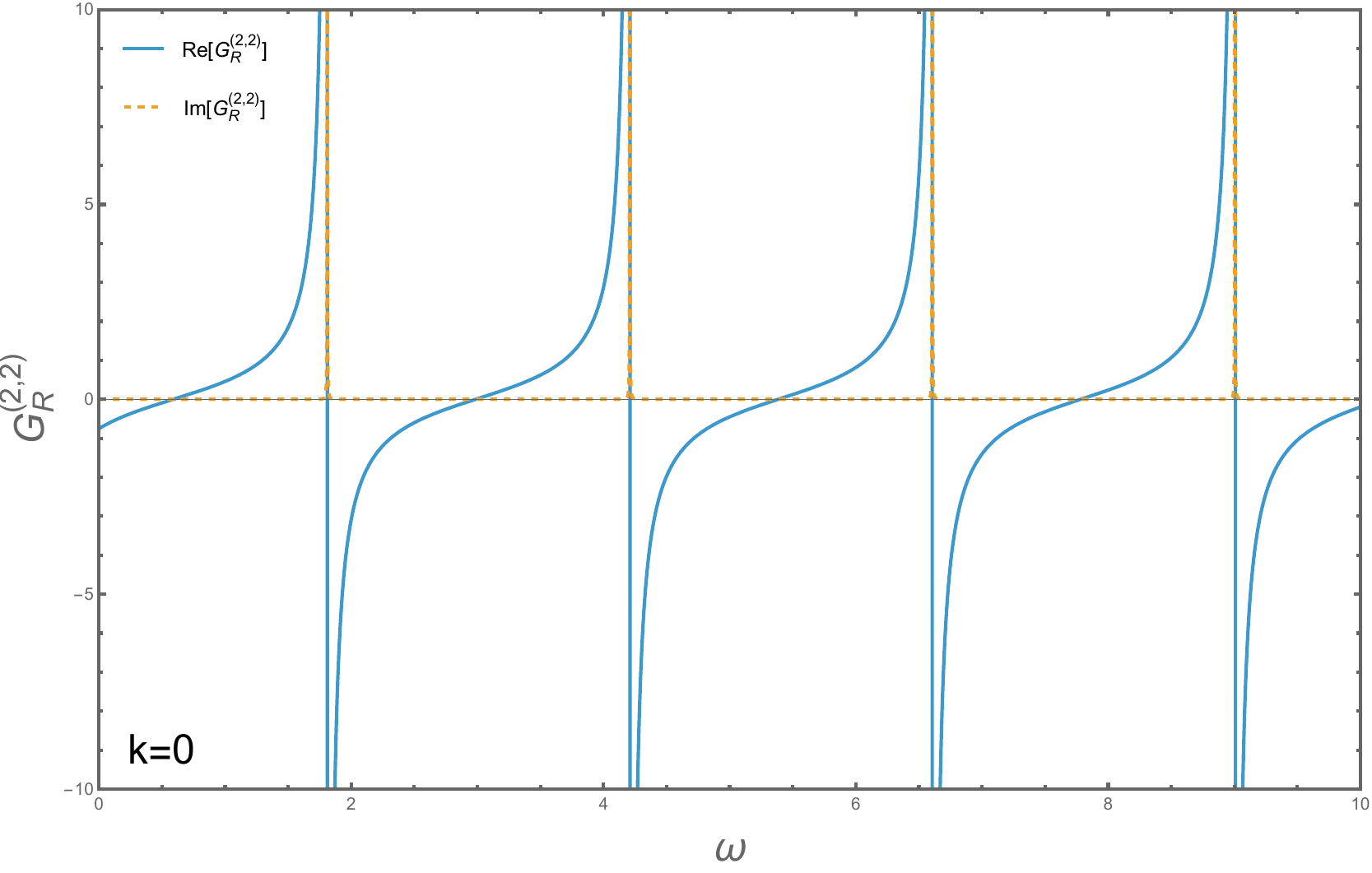}
\par\end{centering}
\caption{\label{fig:8}The holographic correlation function as a function of
$\omega$ at $\mathrm{k}=0$ in confined phase with $q=1$.}
\end{figure}
As a parallel method to investigate the fermionic spectrum, we also
evaluate numerically the equations in (\ref{eq:51}) with respect
to the confined geometry given in (\ref{eq:15}). So the associated
correlation functions are illustrated in Figure \ref{fig:7} and \ref{fig:8}.
Comparing the confined correlation functions with the mass spectrum
given in Table \ref{tab:1} and Figure \ref{fig:1}, we can see several
separated peaks in the correlation function. Since the two-point fermionic
correlation function is proportional to $\left(\cancel{k}-m\right)^{-1}$,
the peaks refer to the separated dispersion curves regarding the various
onset masses. In the confined correlation function, the onset position
of $\omega$ at $\mathrm{k}=0$ agrees with the fermionic spectrum
and it basically holds for nonzero $q$, therefore the holographic
correlation function indicates consistent conclusion of the fermionic
spectrum with the method used in Section 3. 

\section{Summary}

In this work, we study the spectroscopy of baryonic fermion in the
D(-1)-D3 brane system. The geometric background of this system given
by IIB supergravity includes a black and a bubble solution which can
describe respectively a deconfined and a confined gauge theory with
instantons in holography. To simplify the setup, we briefly reduce
the 10d supergravity background to an 5d background of gravity-dilaton-axion
system. By considering a probe spinor in the 5d bulk, we illustrate
the decomposition and dimensional reduction of the spinor, then obtain
the fermionic spectrum in the confined geometry. Since the fermion
must be a gauge-invariant operator according to the gauge-gravity
duality, we further identify the fermion with a baryon in the confined
geometry and compare our mass spectrum with the baryonic experimental
data. However, as most works on holography \cite{key-38,key-39},
the gauge-invariant fermion in the deconfined phase is usually identified
with the baryonic plasmino instead of the baryon since quarks and
gluons can not form a baryon in the deconfined phase. Afterwards,
we derive the equations for the holographic correlation function from
the Dirac equation, then demonstrate the numerical calculations with
the infalling boundary condition both in the deconfined and confined
background. In particular, we expand the correlation function as zero-th
order solution plus first order solution in the deconfined geometry.
Our numerical results in the deconfined case illustrate that the first
order correction to the total correlation function including two branches
of the fermionic dispersion curves is dominated for small momentum,
while the zero-th order correlation function dominates the behavior
of the fermionic dispersion curves at large momentum giving only one
merged dispersion curves on light-cone. This behavior is out of reach
without the presence of the instanton and it is very close to the
fermionic dispersion curves obtained by HTL. Furthermore, we also
find the effective mass generated by the medium effect of fermion
splits into two values which may be due to the spin-dependent interactions
among fermions induced by instantons. Although it has be discussed
qualitatively in the framework of QCD with instantons \cite{key-1,key-2},
it is interesting that the holographic approach reproduces this result.
Besides, in the confined case, the holographic correlation function
explicitly indicates that the position of the onset mass is consistent
with the fermionic spectrum. Altogether, it means seemingly that the
instantonic configuration in QCD is very influential on the spectroscopy
of the fermion.

\section*{Acknowledgements}

We would like to thank Prof.Lin Shu for helpful discussion. This work
is supported by the National Natural Science Foundation of China (NSFC)
under Grant No. 12005033 and the Fundamental Research Funds for the
Central Universities under Grant No. 3132025200.

\end{document}